\documentclass[twocolumn,showpacs,preprintnumbers,amsmath,amssymb,superscriptaddress]{revtex4-2}

\usepackage{graphicx}    
\usepackage{latexsym}    %
\usepackage{color}       %
\usepackage{dcolumn}     
\usepackage{bm}          
\usepackage{amssymb}     
\usepackage{hyperref}    
\expandafter\let\csname equation*\endcsname\relax
\expandafter\let\csname endequation*\endcsname\relax
\usepackage{amsmath}
\usepackage{mathtools}          

\usepackage{graphics}
\usepackage{epsfig}
\usepackage{subfigure}
\usepackage{nicefrac}
\usepackage{verbatim}
\usepackage{enumerate}
\usepackage{footnote}
\usepackage{booktabs}
\usepackage{siunitx}

\def\url#1{}
\def\urlprefix{}

\begin{document}

\title{Explicit construction of joint multipoint statistics in complex systems\\
}
\author{J.~Friedrich}
\affiliation{ForWind, Institute of Physics, University of Oldenburg
Küpkersweg 70, D-26129 Oldenburg, Germany}
\affiliation{Univ. Lyon, ENS de Lyon, Univ. Claude Bernard, CNRS, Laboratoire de Physique,
F-69342, Lyon, France}
\author{J.~Peinke}
\affiliation{ForWind, Institute of Physics, University of Oldenburg
Küpkersweg 70, D-26129 Oldenburg, Germany}
\author{A.~Pumir}
\affiliation{Univ. Lyon, ENS de Lyon, Univ. Claude Bernard, CNRS, Laboratoire de Physique,
F-69342, Lyon, France}
\author{R.~Grauer}
\affiliation{Institute for Theoretical Physics I, Ruhr-University Bochum, Universit\"{a}tsstr. 150,
D-44801 Bochum, Germany}
\date{\today}

\begin{abstract}
Complex systems often involve random fluctuations for which self-similar properties in space and time play an important role.
Fractional Brownian motions, characterized by a single scaling exponent, the Hurst exponent $H$, provide a convenient tool to construct synthetic signals that capture the statistical properties of many processes in the physical sciences and beyond. However, in certain strongly interacting systems, e.g., turbulent flows, stock market indices, or cardiac interbeats, multiscale interactions lead to significant deviations from self-similarity and may therefore require a more elaborate description. In the context of turbulence, the Kolmogorov-Oboukhov model (K62) describes anomalous scaling, albeit explicit constructions of a turbulent signal by this model are not available yet. Here, we
derive an explicit formula for the joint multipoint probability density function of a multifractal field. To this end, we consider a scale mixture of fractional Ornstein-Uhlenbeck processes and introduce a fluctuating length scale in the corresponding covariance function.
In deriving the complete statistical properties of the field, we are able to
systematically model synthetic multifractal phenomena.
We conclude by giving a brief outlook on potential applications which range from specific tailoring or stochastic interpolation of wind fields to the modeling of financial data or non-Gaussian features in geophysical or geospatial settings.
\end{abstract}

\maketitle

\section{Introduction}
The theory of fractals has provided a unifying view on processes that occur in complex systems~\cite{mandelbrot1982fractal}. In particular,
the concept of self-similarity has numerous applications, ranging from the solutions of ordinary and partial differential equations to the description of complex processes in nature~\cite{barenblatt1996scaling,barenblatt1972self,goldenfeld2018lectures}. Typical examples include subsurface hydrology~\cite{molz1997fractional,molz1993fractal}, turbulence measurements in the solar wind~\cite{Goldstein1995}, random amoeboid motion~\cite{Makarava2014}, and heart interbeat fluctuations~\cite{Peng1993}. In this context, it is convenient to describe fluctuating time series
or spatial fields as fractional Brownian motion (fBm) where the roughness of the signal is determined by a single scaling exponent, the Hurst exponent $H$~\cite{levy1965processus,mandelbrot1968,grebenkov2015multiscale}.
Nonetheless, many strongly interacting systems exhibit deviations from  statistical self-similarity. Perhaps one of the most emblematic examples for such multifractal properties is the phenomenon of intermittency in turbulence~\cite{frisch:1995}. The latter manifests itself by the occurence of strong velocity fluctuations at small scales which follow a non-Gaussian distribution. Similar phenomena can also be encountered in other systems, e.g., stock market fluctuations~\cite{ghashghaie1996turbulent}, hydraulic conductivity measurements in geophysics~\cite{https://doi.org/10.1029/2003GL019320}, or urban rent price fields~\cite{HU2012161,https://doi.org/10.17185/duepublico/73834}. Hence, in contrast to fractional Brownian motion, which is a Gaussian process and thus fully determined by its covariance function, multifractal features are inherently non-Gaussian and, therefore, fundamentally more difficult to describe. Accordingly, numerous scholars have been focusing on random multifractal models, which include multiplicative or hierarchical cascade processes~\cite{juneja1994synthetic,malara2016fast,rosales2008anomalous,schertzer1988multifractal} and random multifractal walks~\cite{bacry2001multifractal} (see also recent works in Refs.~\cite{chevillard2010stochastic,Chevillard_2019,chevillard2020multifractal,apolinario2020shot}). The corresponding multifractal models are oftentimes formulated as non-Gaussian generalizations of fBm but a statistical characterization of their moments or probability distribution functions is rather difficult to assess.

Furthermore, the multifractal description has been devised in order to infer the scaling behavior of moments of certain quantities, such as velocity or temperature increments and is - with a few exceptions~\cite{meneveau1990joint,Jaffard_2019} - restricted to a single-scale separation or two-point analysis~\cite{frisch:1995}.
Nonetheless, to explore the spatial and temporal structures of the fluctuating field in more detail, it is useful to go beyond considerations of mere scaling aspects. This can be done by determining the probability of field fluctuations at several points $\mathbf{x}_i$~\cite{monin}.
Hence, the most general quantity in such a statistical description of a random field $\mathbf{u}(\mathbf{x})$ (e.g., a turbulent velocity field) is the joint
$n$-point probability density function (PDF)
\begin{equation}
  f_n(\mathbf{u}_1,\mathbf{x}_1;\ldots;\mathbf{u}_n,\mathbf{x}_n) = \left \langle
  \prod_{i=1}^n \delta(\mathbf{u}_i-\mathbf{u}(\mathbf{x}_i)) \right \rangle\;,
  \label{eq:n-point}
\end{equation}
where brackets denote ensemble averages. The knowledge of the joint
$n$-point PDF is particularly useful to unravel the complex behavior resulting from the interplay between spatial scales and is needed in many applications such as time series reconstructions~\cite{nawroth-peinke:2006,Friedrich1997,Friedrich2011a,Friedrich2017,Sinhuber_2021}, for spherical $n$-point correlations in 2D string theory~\cite{DIFRANCESCO1991385}, many body problems in strongly correlated quantum field theory~\cite{babujian2017multipoint} or solid state physics~\cite{kugler2021multipoint,squarcini2021multipoint,kitanine1999form}, and statistical
turbulence theory~\cite{Lundgren1967,Friedrich2017,mydlarski1998structures,yang2020dynamics}.
Nevertheless, explicit determination of the multipoint PDF (\ref{eq:n-point})
is rather difficult if the field $\mathbf{u}(\mathbf{x})$ exhibits
non-Gaussian properties and one typically has to resort to approximations, e.g., the assumption of a Markov process in scale which reduces Eq. (\ref{eq:n-point}) to a product of three-point quantities~\cite{Friedrich1997,Friedrich2011a,Stresing_2010}.

We introduce here an explicit construction of the joint multipoint PDF (\ref{eq:n-point}) with
multifractal scaling that belongs to the class of the Kolmogorov-Oboukhov (K62) model of turbulence~\cite{kolmogorov:1962,obukhov:1962}. The method is based on an ensemble of fractional Ornstein-Uhlenbeck processes which have been modified by the introduction of fluctuating length scales. It can thus be considered as a multipoint generalization of the two-point statistics proposed by Kolmogorov and Oboukhov~\cite{kolmogorov:1962,obukhov:1962}. In the context of the present work, we notice that the K62 model belongs to the class of ``superstatistics'' put forth by Beck~\cite{BECK2004195}, Castaing~\cite{CASTAING1990177,chabaud1994transition,NAERT199873} and Yakhot~\cite{YAKHOT2006166}.
The concept of superstatistics has also been invoked to describe non-Gaussian behavior, e.g., the anomalous diffusion of nanospheres in an F-actin network~\cite{Wang_2012},
the spreading process of malignant cancer cells~\cite{LEONCHEN20083162}, or the modeling of financial data~\cite{ausloos2003dynamical} (see also~\cite{metzler2020superstatistics} for further reviews on superstatistics).
The main novelty of our method is that it allows for a full statistical characterization (i.e., the determination of the $n$-point PDF (\ref{eq:n-point})) of the multifractal field. Moreover, individual field realizations can be drawn from a so-called Gaussian scale mixture which is thus particularly suitable for modeling purposes, e.g., in the context of mesoscale atmospheric modeling~\cite{schertzer1988multifractal} or synthetic wind field modeling~\cite{https://doi.org/10.1002/we.422}.

The paper is organized as follows: Sec.~\ref{sec:frac_ou} discusses the ensemble of fractional Ornstein-Uhlenbeck processes and derives their respective correlation functions. Furthermore, it is shown through the example of a turbulent velocity field that a lognormal distribution of the fluctuating scale parameters reproduces the Komogorov-Oboukhov scaling of turbulence. In Sec.~\ref{sec:multi} we introduce the multipoint statistics of the fractional Ornstein-Uhlenbeck scale mixture and
explicitly consider one-, two-, and three-point statistical quantities. These quantities are also compared to empirical findings such as the PDF of velocity multipliers~\cite{chen2003kolmogorov}. Further extensions of our method, such as the generalization to a three-dimensional field, are outlined in Sec.~\ref{sec:3d}. Sec.~\ref{sec:outlook} briefly discusses potential applications of the proposed multi-point statistics in the fields of wind energy, geophysics, and also as a potential ansatz for a multi-point hierarchy in turbulence~\cite{Lundgren1967}.

\section{The Kolmogorov-Obukhov model of turbulence as a mixture of fractional Ornstein-Uhlenbeck processes}
\label{sec:frac_ou}
The methodology that is outlined in this section borrows from the superstatistical approach, which describes non-equilibrium systems in terms of a superposition of equilibrium statistics~\cite{BECK2003267}. Hence, we propose a statistical description of a turbulent velocity field $u(x)$ as an ensemble of Gaussian distributed velocity fields $w_{\xi}(x)$. To this end, we use a functional formulation of the velocity field statistics, which was first introduced by Hopf~\cite{Hopf1952}.
In the following, we restrict ourselves to the case of a one-dimensional field $u(x)$ whereas a generalization to multiple dimensions is discussed in Sec.~\ref{sec:3d}.
We are thus led to consider the characteristic functional
\begin{equation}
  \varphi[\alpha] = \left \langle e^{i \int \textrm{d}x \alpha(x)u(x)}\right \rangle\;,
  \label{eq:charac}
\end{equation}
in terms of a superposition of characteristic functionals $\varphi_\xi[\alpha]$ belonging to Gaussian sub-ensembles
\begin{equation}
  \varphi[\alpha] = \int \textrm{d}\xi g(\xi)\varphi_\xi[\alpha]\;.
  \label{eq:charac_lam}
\end{equation}
Here, $g(\xi)$ denotes the PDF of a parameter $\xi$ that is arbitrary at this stage. Due to the Gaussianity of the sub-ensemble, the characteristic functional can be calculated explicitly (see for instance~\cite{monin,Wilczek_2016,10.1007/978-3-319-57934-4_4}) as
\begin{equation}
  \varphi_\xi[\alpha]= e^{-\frac{1}{2}\int \textrm{d}x \int \textrm{d}x' \alpha(x)C_\xi(x-x')\alpha(x')}\;,
  \label{eq:charac_gauss}
\end{equation}
where we assume that the sub-ensemble possesses no mean and is thus fully characterized by the correlation function $C_{\xi}(r) =\langle w_\xi(x+r)w_\xi(x) \rangle$ of a Gaussian random field $w_\xi(x)$. Furthermore, we make use of the assumption of homogeneity which entails that the two-point correlation function is a function of the spatial separation $r$ only.

In the next step, we determine the correlation function $C_{\xi}(r)$. To this end,
we consider a fractional Ornstein-Uhlenbeck process $w(x)$ which obeys the following ordinary stochastic differential equation
\begin{equation}
 \textrm{d}w(x) = -\frac{1}{L} w(x) \textrm{d}x + \textrm{d}B^{H}(x)\;,
 \label{eq:fOU}
\end{equation}
where $\textrm{d}B^{H}(x)$ denotes the increment of fractional Gaussian noise with covariance
\begin{equation}
 \left \langle B^{H}(x) B^{H}(x') \right \rangle = \frac{\sigma^2}{2} \left(|x|^{2H}+|x'|^{2H}-|x-x'|^{2H} \right)\;.
\end{equation}
Here, $H$ is the Hurst exponent which lies in between $0$ and $1$.
The explicit calculation of the correlation function $C(r) =\langle w(x+r)w(x) \rangle$ of the stationary fractional Ornstein-Uhlenbeck process can be found in
Mardoukhi et al.~\cite{Mardoukhi_2020} and is reproduced in Appendix~\ref{app:fOU}.
In order to introduce fluctuations at multiple points of the scale mixture (\ref{eq:charac_lam}),
we replace the scale $r$ in the correlation function $C(r)$ by a parametrized scale $\rho_\xi(r)$.
Hence, we consider a family of Gaussian processes $w_{\xi}(x)$ with zero mean, which are characterized by the correlation function
\begin{widetext}
\begin{align}
  C_{\xi}(r)=-\frac{\sigma^2}{2} \rho_\xi(r)^{2H}  +\frac{\sigma^2 L^{2H}}{2}\Gamma(2H+1)\cosh \frac{\rho_\xi(r)}{L}
  + \frac{\sigma^2 \rho_\xi(r)^{2H+1} }{4L(2H+1)}  \left[
  e^{-\frac{\rho_\xi(r)}{L}} \mathcal{K}^H  \left(\frac{\rho_\xi(r)}{L} \right) -e^{\frac{\rho_\xi(r)}{L}} \mathcal{K}^H  \left(-\frac{\rho_\xi(r)}{L} \right) \right]\;,
  \label{eq:corr_lambda}
\end{align}
\end{widetext}
where we imply that $r \ge0$. Furthermore, we introduce Kummer's confluent hypergeometric function
\begin{equation}
   {_1}F_1 (a,b,z)= \frac{\Gamma(b)}{\Gamma(b-a)\Gamma(a)} \int_0^1 \textrm{d}t e^{zt}t^{a-1}(1-t)^{b-a-1}\;,
\end{equation}
as $\mathcal{K}^H\left(r \right)={_1}F_1 (2H+1,2H+2,r)$.

We thus parametrize the correlation function of the fractional Ornstein-Uhlenbeck process by allowing for a more general function of the scale separation $r$, namely $\rho_\xi(r)$. In the following, we consider an explicit form of this generalized scale
\begin{equation}
  \rho_\xi(r)= \xi^{\sqrt{A+\mu \ln_+ \frac{x^{\succ}}{r+x^{\prec}}}}\left(\frac{r+x^{\prec}}{x^{\succ}}\right)^{\frac{\mu}{2}}r\;,
  \label{eq:rho_lambda}
\end{equation}
where $\mu$ denotes the intermittency coefficient, $x^{\prec}$ a small-scale cut-off, and $x^{\succ}$ a large-scale quantity which does not necessarily have to coincide with the integral length scale $L$.
Hence, by this re-parametrization of the scale $r$ to $\rho_\xi(r)$, we have introduced a family of Gaussian processes with two-point correlations that are a varying function of the parameter $\xi$ and thus deviate from the simple inertial range scaling $C(r) \sim r^{2H}$ of the ordinary fractional Ornstein-Uhlenbeck process $w(r)$.
Moreover, the logarithm in Eq. (\ref{eq:rho_lambda}) is restricted to positive values only
\begin{equation}
  \ln_+ x = \begin{cases}
  \ln x & \textrm{for}\quad  x\ge 1\;,\\
  0 & \textrm{for} \quad 0<x<1\;.
\end{cases}
\end{equation}
It should be noted that for $\mu=0$ and $A=0$, $\rho_\xi(x)=r$, and the correlation function (\ref{eq:corr_lambda}) reduces to the (stationary)  correlation function of the fractional Ornstein-Uhlenbeck process~\cite{Mardoukhi_2020}.
Furthermore, in the limit $r \rightarrow \infty$, the asymptotic behavior of Kummer's confluent hypergeometric function suggests
${_1}F_1(a,b,z) \sim \Gamma(b) \left[\frac{e^z z^{a-b}}{\Gamma(a)} + \frac{(-z)^{-a}}{\Gamma(b-a)} \right]$
and we obtain $\lim_{r \rightarrow \infty} C_\xi(r)=0$.

In the final step, we demand that the parameter $\xi$ follows a lognormal distribution
\begin{equation}
  g(\xi) = \frac{1}{\sqrt{2 \pi}\xi}e^{-\frac{[\ln \xi]^2}{2}}\;,
  \label{eq:lognormal}
\end{equation}
which results in a scale mixture (\ref{eq:charac_lam}) that exhibits strong correlations between multiple points.

We will now verify that the increment statistics derived from (\ref{eq:charac_lam}) belongs to the class of the K62 model of turbulence. To this end, we calculate the velocity increment PDF
\begin{align}\nonumber
 \lefteqn{h(v,r)=\left \langle \delta(v-u(x+r)+u(x)) \right \rangle}\\ \nonumber
 =& \frac{1}{2 \pi}\int \textrm{d}w e^{-iwv} \left \langle e^{iw[u(x+r)-u(x)]} \right \rangle \\ \nonumber
 =& \frac{1}{2 \pi}\int \textrm{d}w e^{-iwv} \varphi \left[\alpha(x')=w\{\delta(x'-x-r)-\delta(x'-x)\}\right]\\
 =& \frac{1}{2 \pi}\int \textrm{d}w e^{-iwv}
 \int_0^{\infty} \textrm{d}\xi g(\xi) e^{-\frac{w^2}{2}\left[2C_\xi(0)-2C_\xi(r) \right]}\;.
\end{align}
Here, we can introduce the second-order structure function
\begin{equation}
 S_{2,\xi}(r)=2C_\xi(0)-2C_\xi(r)
 \label{eq:s2_lambda}
\end{equation} 
%
%
which yields
\begin{equation}
  h(v,r)= \int_0^{\infty} \textrm{d}\xi g(\xi)
  \frac{1}{\sqrt{2 \pi S_{2,\xi}(r)}}
  \exp \left[-\frac{v^2}{2S_{2,\xi}(r)} \right]\;.
  \label{eq:inc_pdf}
\end{equation}
Taking the moments of $h(v,r)$ yields
\begin{align}\nonumber
  \lefteqn{\left \langle [u(x+r)-u(x)]^n \right \rangle= \int \textrm{d}v v^n h(v,r)}\\
  =& \int_0^{\infty} \textrm{d}\xi g(\xi)
   (n-1)!! [S_{2,\xi}(r)]^{n/2}\;,
   \label{eq:moments}
\end{align}
and for large $L$, we can approximate $S_{2,\xi}(r)\approx \sigma^2\rho_\xi(r)^{2H}$ and obtain
\begin{align}\nonumber
  \lefteqn{\left \langle [u(x+r)-u(x)]^n \right \rangle}\\ \nonumber
   =& (n-1)!! \sigma^{n} \small{ \left(\frac{r+x^\prec}{x^{\succ}}\right)^{\frac{\mu}{2}n H}} \hspace{-0.18cm}r^{nH}\underbrace{\int \hspace{-0.12cm} \textrm{d}\xi g(\xi)\xi^{nH\sqrt{A+\mu \ln_+ \frac{x^\succ}{r+x^\prec}}}}_{=e^{-\frac{1}{2}n^2 H^2
   \left(A+\mu \ln_+ \frac{x^\succ}{r+x^\prec} \right)}} \\
   =& C_n
   \left(\frac{r+x^\prec}{x^\succ}\right)^{\frac{\mu}{2}(nH- n^2H^2)}r^{nH}\;,
   \label{eq:k62}
\end{align}
%
which, for $H=1/3$ and in the limit of $r \gg x^{\prec}$, reduces to the original K62 scaling with scaling exponents $\zeta_n
= \frac{n}{3}-\frac{\mu}{18}n(n-3)$.
Moreover, for $r \ll L$, it is possible to transform the one-increment PDF (\ref{eq:inc_pdf})  into a form similar to the PDF proposed by Yakhot~\cite{YAKHOT2006166} which can be obtained from a Mellin transform of the structure functions
(\ref{eq:k62}). To this end, we substitute $\xi'=\frac{v}{\sigma \rho_\xi(r)^{H}}$ for $\xi$ and obtain
%
\begin{equation}
 h(v,r) = \int_0^{\infty} \textrm{d}\xi' e^{-\frac{\xi'^2}{2}} \frac{\exp \left[- \frac{\left(\ln \frac{v}{\sigma \xi' \left(\frac{r+x^\prec}{x^\succ}\right)^{\frac{\mu H}{2}}r^H}
 \right)^2}{2H^2 \left(A+\mu \ln_+ \frac{x^\succ}{r+x^\prec}\right)} \right]}{2\pi v H \sqrt{A+\mu \ln_+ \frac{x^\succ}{r+x^\prec}}} \;.
 \label{eq:inc_pdf_yakhot}
\end{equation}
%
Similar formulas for the increment PDF have also been proposed by Beck~\cite{BECK2004195} as well as by Castaing~\cite{CASTAING1990177} who allowed for a more general form of Eq. (\ref{eq:inc_pdf_yakhot}) which does not necessarily entail scaling solutions (\ref{eq:k62}).

We want to end this section by a recapitulation of the three steps that leed to the reconstruction of the characteristic functional (\ref{eq:charac_lam}) which contains the K62 increment statistics in Eqs. (\ref{eq:k62}) and (\ref{eq:inc_pdf_yakhot}):
\emph{i.)} we consider a Gaussian random process in form of a fractional Ornstein-Uhlenbeck process $w(r)$ whose monofractal behavior is  characterized by the Hurst exponent $H$ and whose two-point correlation function is given by $C(r)$, \emph{ii.)} in this correlation function, we introduce a re-parameterization of the scale parameter $r \rightarrow \rho_{\xi}(r)$. We thus consider a family of Gaussian processes with fluctuating two-point correlation functions (\ref{eq:corr_lambda}). We propose an explicit form for the re-parametrization in Eq. (\ref{eq:rho_lambda}), and \emph{iii.)} we sum over all fluctuations of the parameter $\xi$ by weighing it according to a lognormal distribution (\ref{eq:lognormal}). Hence, by virtue of these three step process, we introduce strong correlations which culminate in the non-Gaussian statistics of field increments (\ref{eq:inc_pdf_yakhot}).
This will be further elaborated upon in the following section, where we derive the main result of this paper, i.e., the multipoint statistics of the Kolmogorov-Oboukhov-type velocity field.
\section{Multipoint statistics of the Gaussian scale mixture}
\label{sec:multi}
\begin{figure}
  \includegraphics[width=0.48\textwidth]{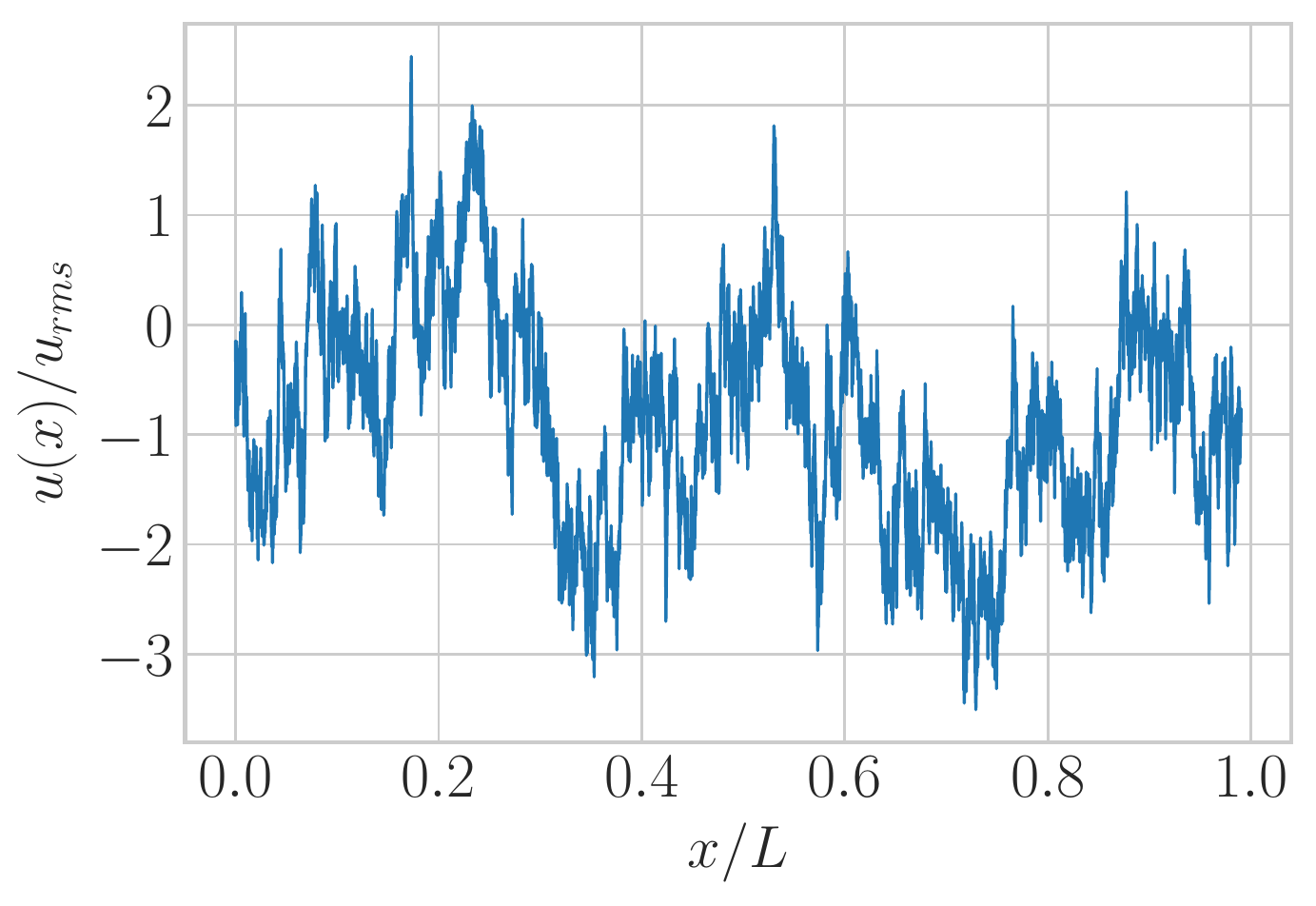}
  \caption{Typical realization of the velocity field $u(x)$ drawn from the compound $n$-point PDF (\ref{eq:ensemble_n_point}) with spatial resolution $n=4096$.}
  \label{fig:u_k62}
\end{figure}

As it has been shown in the previous section, the Gaussian scale mixture (\ref{eq:charac_lam}) with correlation function (\ref{eq:corr_lambda}) and the parameter distribution leads to the K62-scaling of velocity increments (\ref{eq:k62}). In this section, we derive the joint $n$-point probability density function of the ensemble
\begin{equation}
  f_n(u_1,x_1;\ldots;u_n,x_n) = \left \langle
  \prod_{i=1}^n \delta(u_i-u(x_i)) \right \rangle\;.
  \end{equation}
From Eq. (\ref{eq:charac_lam}), we obtain the $n$-point PDF
\begin{equation}
  f_n(u_1,x_1;\ldots;u_n,x_n)
  = \int \textrm{d}\xi g(\xi)
  f_{n,\xi}(u_1,x_1;\ldots;u_n,x_n),
  \label{eq:ensemble_n_point}
\end{equation}
as a superposition of multivariate normal distributions
\begin{equation}
  f_{n,\xi}(u_1,x_1;\ldots;u_n,x_n)=
  \frac{1}{\sqrt{(2\pi)^n \det \sigma_\xi}}e^{-\frac{1}{2}\mathbf{u}^T \sigma_\xi^{-1} \mathbf{u}}\;,
\end{equation}
where the parameter $\xi$ of the covariance matrices $\sigma_{ij,\xi}=
C_\xi(x_i,x_j)$ in Eq. (\ref{eq:corr_lambda}) is distributed according to a lognormal distribution (\ref{eq:lognormal}). Hence, the $n$-point PDF (\ref{eq:ensemble_n_point}) belongs to the class of compound
probability distributions~\cite{feller1957introduction}. Moreover, as the parameter $\xi$ in Eq. (\ref{eq:rho_lambda}) represents a scale parameter, the multipoint model can also be apprehended as a multivariate Gaussian scale mixture which, in this particular case, leads to heavy-tailed behavior
(see Eq. (\ref{eq:inc_pdf_yakhot})). In the following, we want to further examine the multipoint statistics governed by Eq. (\ref{eq:ensemble_n_point}).
In the general case (i.e., for $n>1$), Eq. (\ref{eq:ensemble_n_point}) has no closed form and one has to resort to numerical treatments such as Markov Chain Monte Carlo or collapsed Gibbs sampling. In the following, we draw samples from the $n$-point PDF (\ref{eq:ensemble_n_point}) by means of a multivariate Gaussian mixture model using TensorFlow~\cite{dillon2017tensorflow}. A typical realization is depicted in Fig.~\ref{fig:u_k62} with model parameters
$L=l=1.01$, $\eta=0.0005$, $H=1/3$, $\mu=0.227$,
and a spatial resolution of $N=4096$. Furthermore, in order to generate the Gaussian scale mixture, we consider 200 realizations of the lognormally
distributed parameter $\xi$.
In the following sections, we compare the statistics of the simulated velocity field realizations to explicitly derived one-, two-, and three-point quantities of the Gaussian scale mixture.
\begin{figure}
  \includegraphics[width=0.48\textwidth]{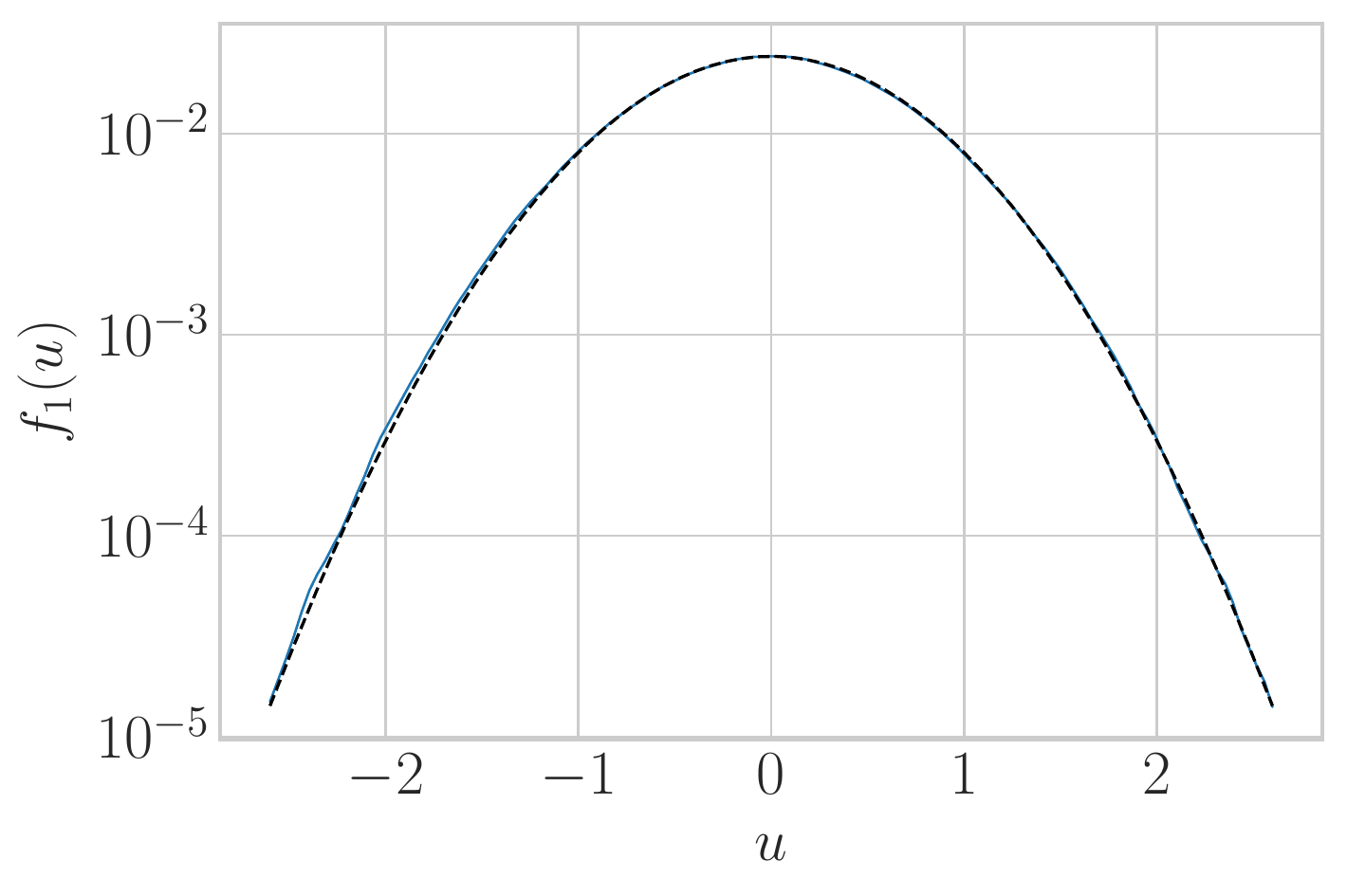}
  \caption{Single-point PDF from 5000 realizations of velocity fields similar to the one depicted in Fig.~\ref{fig:u_k62} Dashed lines correspond to the prediction (\ref{eq:one_point}).}
  \label{fig:single_pdf}
\end{figure}
%
%
\begin{figure}
  \includegraphics[width=0.48\textwidth]{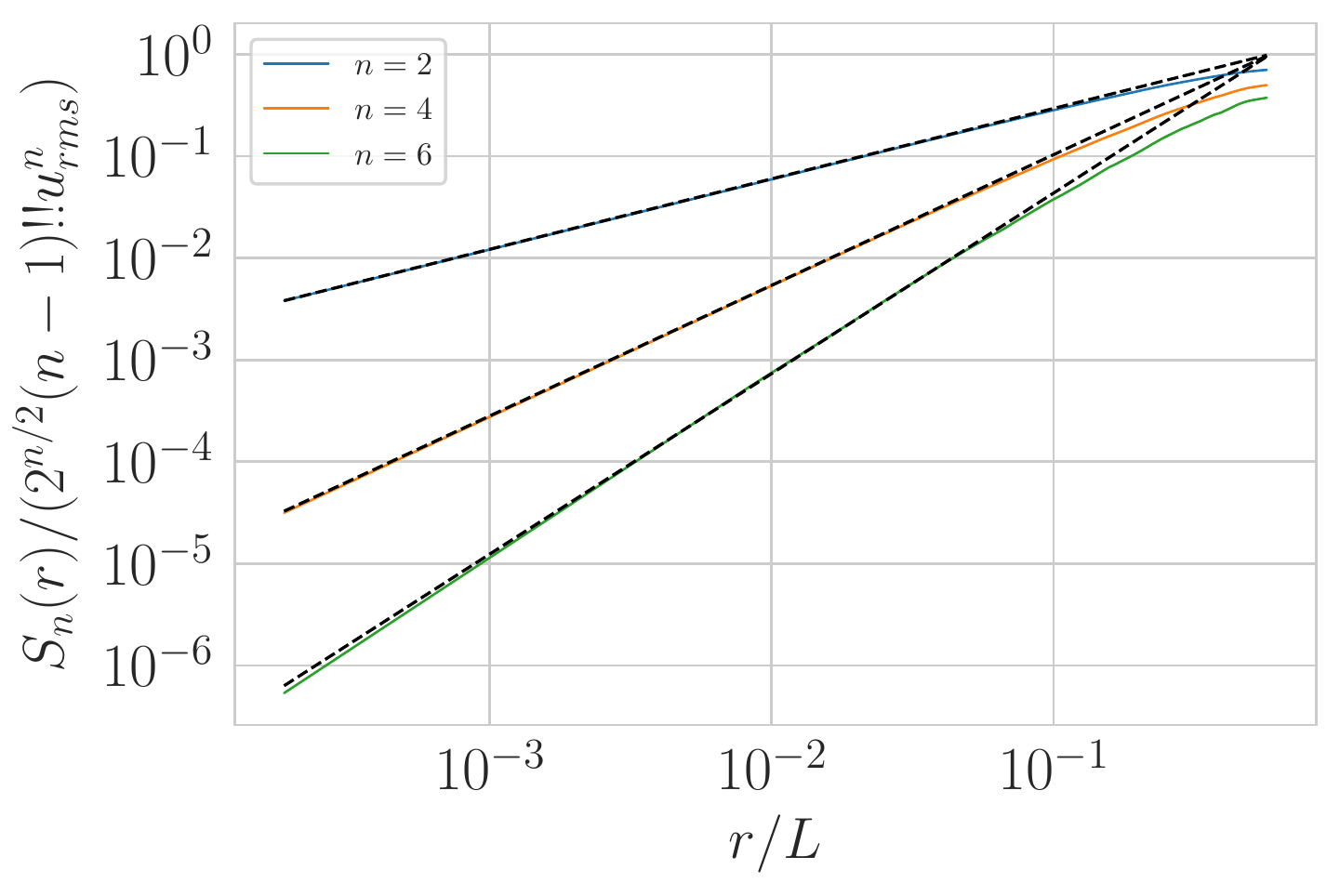}
  \caption{Structure functions of order $n=2,4,6$ determined from 5000 realizations similar to Fig.~\ref{fig:u_k62}. Dashed lines correspond to the prediction (\ref{eq:k62}) of the K62 model.}
  \label{fig:struc}
\end{figure}
\begin{figure}
  \includegraphics[width=0.48\textwidth]{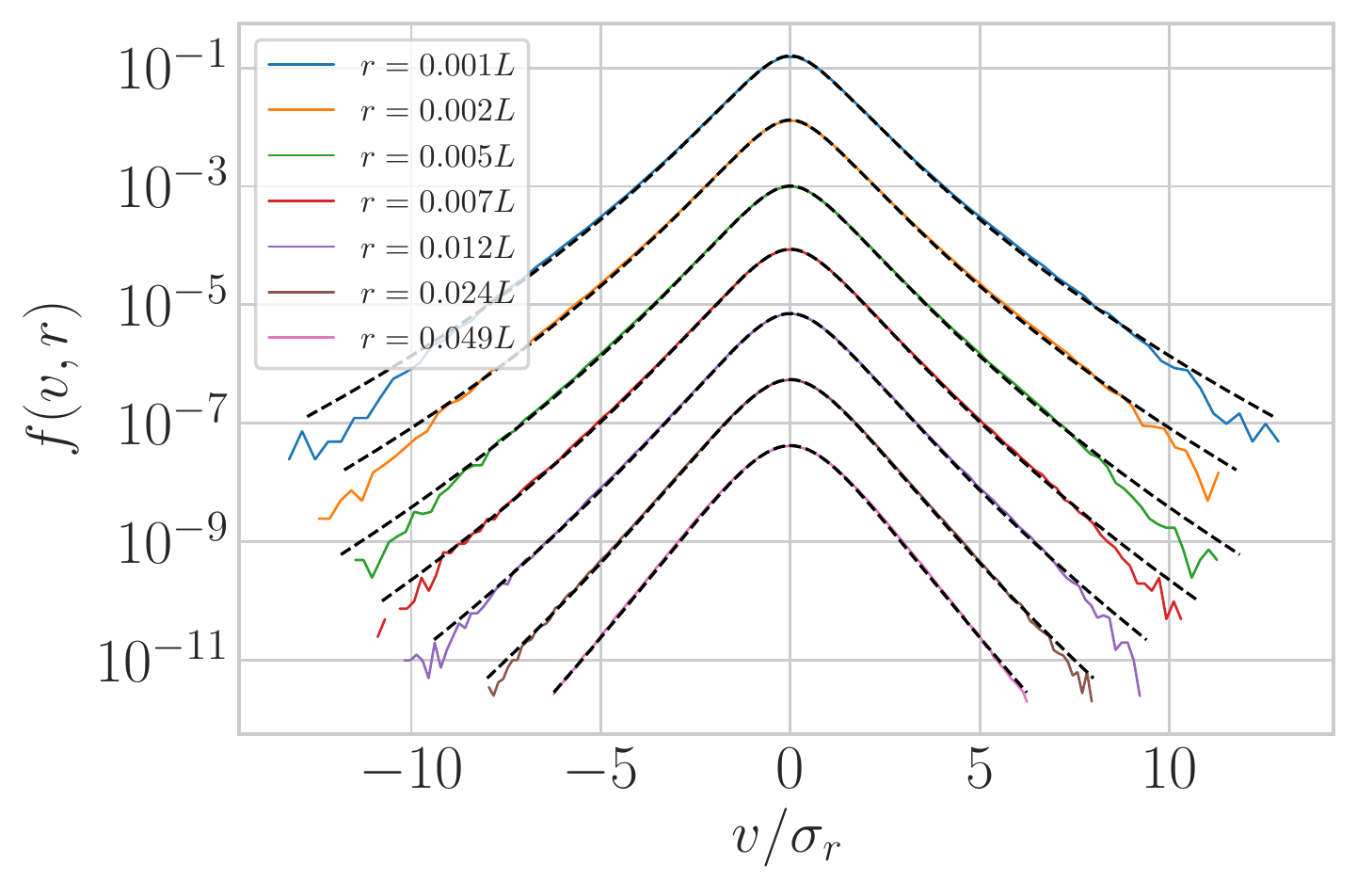}
  \caption{Evolution of the velocity increment PDF determined from 5000 realizations similar to Fig.~\ref{fig:u_k62}. Dashed lines correspond to the prediction (\ref{eq:inc_pdf}).}
  \label{fig:pdf}
\end{figure}
\subsection{Single-point statistical quantities}
\label{sec:one}
Due to the stationarity of the underlying fractional Ornstein-Uhlenbeck process, the single-point PDF $f_1(u_1,x_1)$ becomes independent of $x_1$ and reduces to
\begin{equation}
  f_1(u_1,x_1) = \int \textrm{d}\xi g(\xi)  \frac{1}{\sqrt{2\pi C_\xi(0)}}e^{-\frac{u_1^2}{2 C_\xi(0)}}\;.
\end{equation}
From Eq. (\ref{eq:corr_lambda}) we obtain $C_\xi(0)=\frac{\sigma^2 L^{2H}}{2}\Gamma(2H+1)$ which is independent of the parameter $\xi$ and yields a Gaussian single-point PDF
\begin{equation}
  f_1(u_1) = \frac{1}{\sqrt{\pi \Gamma(2H+1)}\sigma L^H }e^{-\frac{u_1^2}{ \sigma^2 L^{2H} \Gamma(2H+1)}}\;.
  \label{eq:one_point}
\end{equation}
Hence, the correlation function (\ref{eq:corr_lambda}) of the fractional Ornstein-Uhlenbeck process
ensures that the multipoint statistics is homogeneous.  

Fig.~\ref{fig:single_pdf} depicts the single-point PDF (blue) obtained from 5000 realizations similar to the one in Fig.~\ref{fig:u_k62}. The dashed line corresponds to Eq. (\ref{eq:one_point}) and agrees well with the numerical simulations. Thhe current framework thus entails a Gaussian single-point velocity PDF and non-Gaussian features emerge in higher-order statistics such as the two-point quantities discussed in the next section.

\subsection{Two-point statistical quantities}
\label{sec:two}
The correlation function $C(r)=\langle u(x+r)u(x)\rangle$ can readily be derived from Eq. (\ref{eq:ensemble_n_point}) and yields
\begin{equation}
  C(r)=\int \textrm{d} \xi g(\xi)C_\xi(r)\;,
\end{equation}
which implies that $C(0)= \frac{\sigma^2 L^{2H} \Gamma(2H+1)}{2}$ and thus coincides with the variance of the individual fractional Ornstein-Uhlenbeck process. By the same token, we can calculate the structure functions of the velocity increments $\delta_r v(x) = u(x+r)-u(x)$
\begin{equation}
  S_n(r)= \left \langle (\delta_r v)^n \right \rangle=\int \textrm{d}\xi g(\xi) (n-1)!! \left [S_{2,\xi}(r) \right]^{n/2}\;,
\end{equation}
which obviously agree with the moments of the increment PDF derived in Eq. (\ref{eq:moments}). The structure functions of the simulated Gaussian scale mixture model for $n=2,4,6$ are depicted in Fig.~\ref{fig:struc} and agree well with the theoretical predictions of the K62 model (dashed lines) at small scales. At large scales the structure functions saturate which is a consequence of the underlying fractional Ornstein-Uhlenbeck processes. Indeed, due to the asymptotic behavior of Kummer's confluent hypergeometric function we obtain $ \lim_{r \rightarrow \infty}S_n(r)= (n-1)!! 2^{n/2}u_{rms}^n$, where $u_{rms}^2=\frac{\sigma^2 L^2 \Gamma(2H+1)}{2}$.

Furthermore, it has to be stressed that all odd-order structure functions are strictly zero which also entails that the multipoint statistics possesses no third-order structure function. This result is inconsistent with the so-called $4/5$-law for the third-order structure function which can be derived from the Navier-Stokes equation and suggests $S_3(r)=-\frac{4}{5}\left \langle \varepsilon \right \rangle r$, where $r$ lies within the inertial range of scales~\cite{frisch:1995}. Here, $\left \langle \varepsilon \right \rangle$ is defined as the average local energy dissipation rate, which for the present one-dimensional velocity field is defined according to
\begin{equation}
  \left \langle \varepsilon \right \rangle= 2
  \nu \left \langle \left( \frac{\partial u(x)}{\partial x} \right)^2 \right \rangle\;.
  \label{eq:eps}
\end{equation}
The statistics of velocity gradients, however, cannot be predicted by the present framework due to the non-differentiability of the velocity field. This is also highlighted by the fact that structure functions in Fig.~\ref{fig:struc} exhibit no dissipative range which would imply that $S_n(r) \sim r^n$ for small $r$.
In Sec.~\ref{sec:outlook}, we will briefly discuss how skewness and differentiability can be incorporated in the current framework.

The velocity increment PDFs $f(v,r)$ are depicted in Fig.~\ref{fig:pdf} and agree quantitatively with the theoretical predictions which have been obtained from a numerical evaluation of Eq. (\ref{eq:inc_pdf}). In particular, the increment PDF exhibits pronounced tails at small scales $r$.
\subsection{Three-point statistical quantities}
\label{sec:three}
As it has been discussed in the previous section, the Gaussian scale mixture was devised in a way to ensure that two-point statistical quantities obey the K62 phenomenology in the inertial range. Therefore, structure functions as well as velocity increment PDFs in Figs.~\ref{fig:struc} and~\ref{fig:pdf} agree with phenomenological predictions.
Hence, in this section, we want to investigate the behavior of three-point statistical quantities. Such quantities are quite important in the context of multiscale models of turbulence, e.g., the operator product expansion/fusion rules~\cite{Eyink1993,Lvov1996} or a stochastic interpretation of the turbulent energy cascade as a Markov process of velocity increments in scale~\cite{Friedrich1997,Friedrich2011a}. In the following, we want to consider the PDF of the ratio of velocity increments
\begin{equation}
  p(w,r,r') = \left \langle \delta\left(w-\frac{\delta_r v(x)}{\delta_{r'} v(x)} \right)\right \rangle\;,
  \label{eq:ratio_pdf}
\end{equation}
for $r < r'$,
which was first discussed by Kolmogorov~\cite{kolmogorov:1962} and is also referred to as the PDF of velocity increment multipliers~\cite{chen2003kolmogorov}. On the basis of experimental data, Chen et al.~\cite{chen2003kolmogorov} argued that the multiplier PDF is reproduced by a Cauchy distribution derived for increments of fractional Brownian motion with Hurst parameter $H$, namely
\begin{equation}
    p_{fBm}(w,r,r') = \frac{1}{\pi}  \frac{\gamma(r,r')}{\left[w-w_{0}(r,r') \right]^2 +\gamma(r,r')^2 }\;,
    \label{eq:cauchy_fbm}
\end{equation}
with scale and location parameter given by
\begin{align}\label{eq:scale}
  \gamma(r,r') =& \sqrt{1-\frac{\left[r^{2H}+r'^{2H}-(r'-r)^{2H}\right]^2}{4r^{2H} r'^{2H}}}\frac{r^H}{r'^H}\;, \\
  w_0(r,r') =& \frac{r^{2H}+r'^{2H}-(r'-r)^{2H}}{2r'^{H}}\;.
  \label{eq:location}
\end{align}
In the following, we want to determine the multiplier PDF of the Gaussian scale mixture and estimate intermittency corrections. It will be shown that the multiplier PDF agrees fairly well with the prediction from fBm (\ref{eq:cauchy_fbm}) and that intermittency corrections are largely dominated by self-similar features (the ratio of two Gaussian distributed random variables follows a Cauchy distribution~\cite{renyi2007foundations}) and thus might not be observable in turbulence experiments or numerical simulations. Similar arguments have already been put forth by Siefert and Peinke~\cite{siefert2007complete} who reproduced the Cauchy distribution by an ordinary Ornstein-Uhlenbeck process.

The multiplier PDF (\ref{eq:ratio_pdf}) can be obtained from the two-increment PDF
\begin{equation}
  h_2(v,r;v',r')= \left \langle \delta(v-\delta_r v(x))\delta(v'-\delta_{r'}v (x)) \right \rangle\;,
  \label{eq:two_inc}
\end{equation}
according to
\begin{equation}
  p(w,r,r') = \int \textrm{d}v \int \textrm{d}v' \delta\left(w-\frac{v}{v'}\right) h_2(v,r;v',r')\;.
  \label{eq:multi_two}
\end{equation}
We thus derive the two-increment PDF from the three-point PDF (Eq. (\ref{eq:ensemble_n_point}) for $n=3$) according to
\begin{align}\nonumber
    \lefteqn{h_2(v,r;v',r')} \\ \nonumber
    =& \int \textrm{d}u_1 \textrm{d}u_2  \textrm{d}u_3\delta(v-u_3+u_1)\delta(v'-u_2+u_1)\\ \nonumber
    ~&
     \int  \textrm{d}x_1\textrm{d}x_2 \textrm{d}x_3 \delta(r-x_3+x_1)\delta(r'-x_2+x_1)\delta(x-x_1) \\
     & f_3(u_1,x_1;u_2,x_2;u_3,x_3)\;,
\end{align}
which results in a compound bivariate normal distribution
\begin{widetext}
\begin{align}
  h_2(v,r;v',r') =\int \textrm{d}\xi g(\xi) \frac{1}{2 \pi \sqrt{S_{2,\xi}(r)S_{2,\xi}(r')} \sqrt{1-\rho(r,r')^2}}\exp \left[-\frac{\frac{v^2}{S_2(r)} -\frac{2 \rho(r,r')v v'}{\sqrt{S_{2,\xi}(r)S_{2,\xi}(r')}}
  + \frac{v^2}{S_{2,\xi}(r)}}{2(1-\rho(r,r')^2)}  \right]\;,
\end{align}
\end{widetext}
where the correlation $\rho(r,r')$ between two increments is defined as
\begin{equation}
  \rho(r,r')=\frac{S_{2,\xi}(r)+S_{2,\xi}(r')-S_{2,\xi}(r'-r)}{2\sqrt{S_{2,\xi}(r)S_{2,\xi}(r')}}\;.
\end{equation}
Inserting the two-increment PDF into Eq. (\ref{eq:multi_two}) yields
\begin{align}
  p(w,r,r')=\int \textrm{d}\xi g(\xi) \frac{1}{\pi}  \frac{\gamma_{\xi}(r,r')}{\left[w-w_{0,\xi}(r,r') \right]^2 +\gamma_{\xi}(r,r')^2 }\;.
  \label{eq:cauchy_compound}
\end{align}
Hence, the multiplier PDF is a compound Cauchy distribution with location and scale parameter defined as
\begin{align}
  \gamma_{\xi}(r,r')^2=& \sqrt{1-\rho(r,r')^2}\sqrt{\frac{S_{2,\xi}(r)}{S_{2,\xi}(r')}}\;, \\
\textrm{and} \quad w_{0,\xi}(r,r') =& \sqrt{\frac{S_{2,\xi}(r)}{S_{2,\xi}(r')}}
  \rho(r,r')\;,
\end{align}
respectively.
\begin{figure}
  \includegraphics[width=0.48\textwidth]{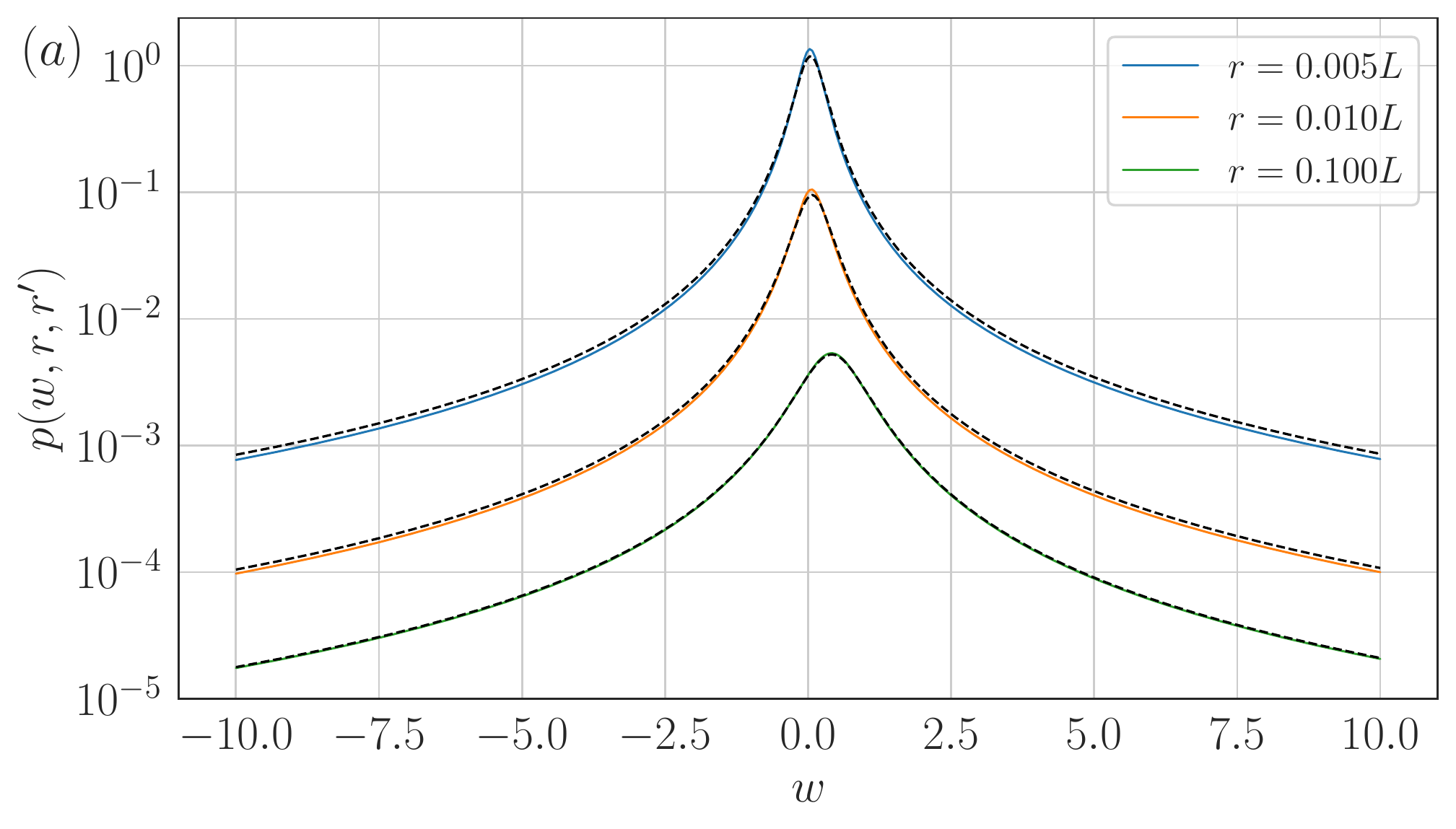}
  \includegraphics[width=0.48\textwidth]{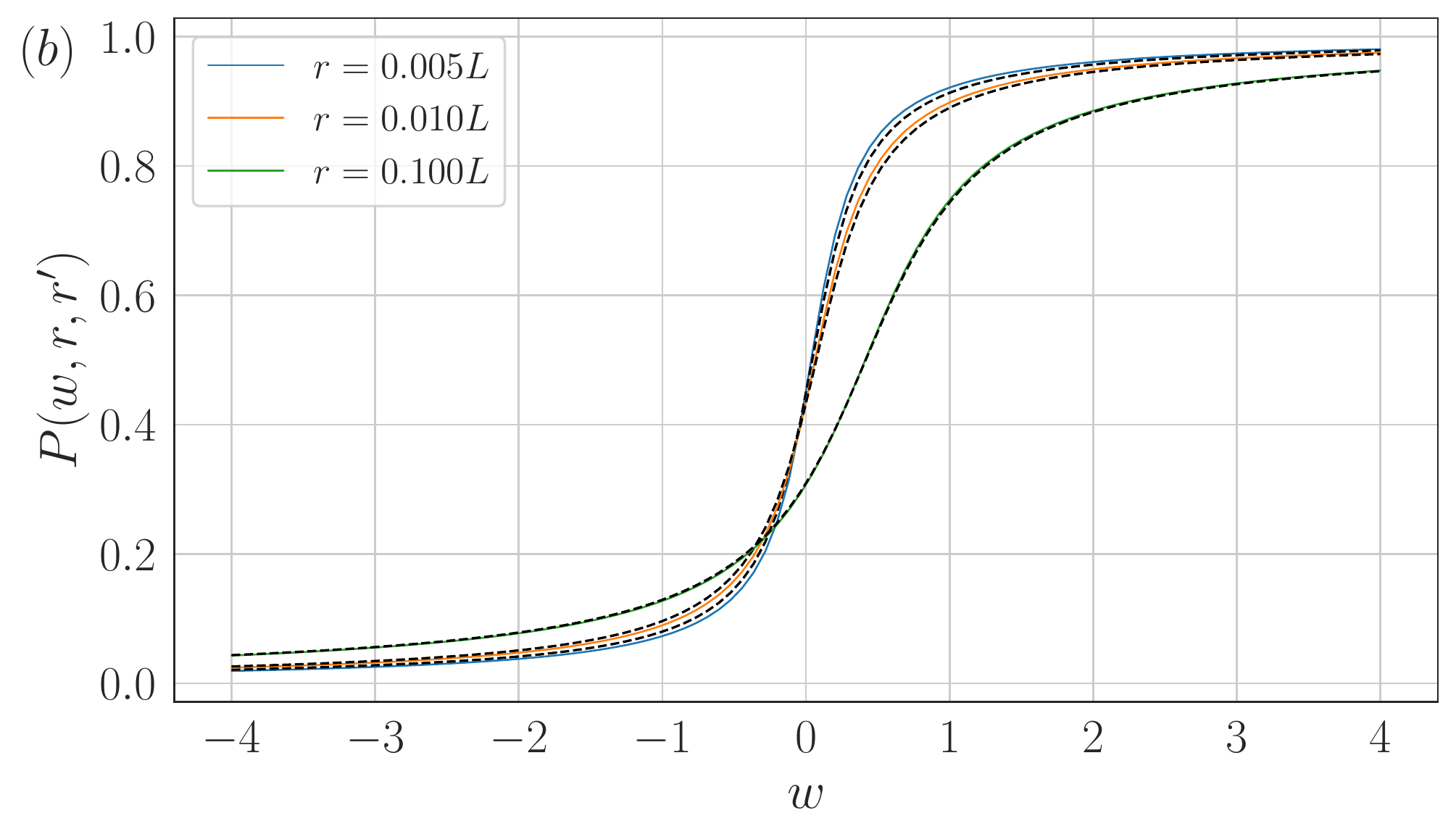}
  \includegraphics[width=0.48\textwidth]{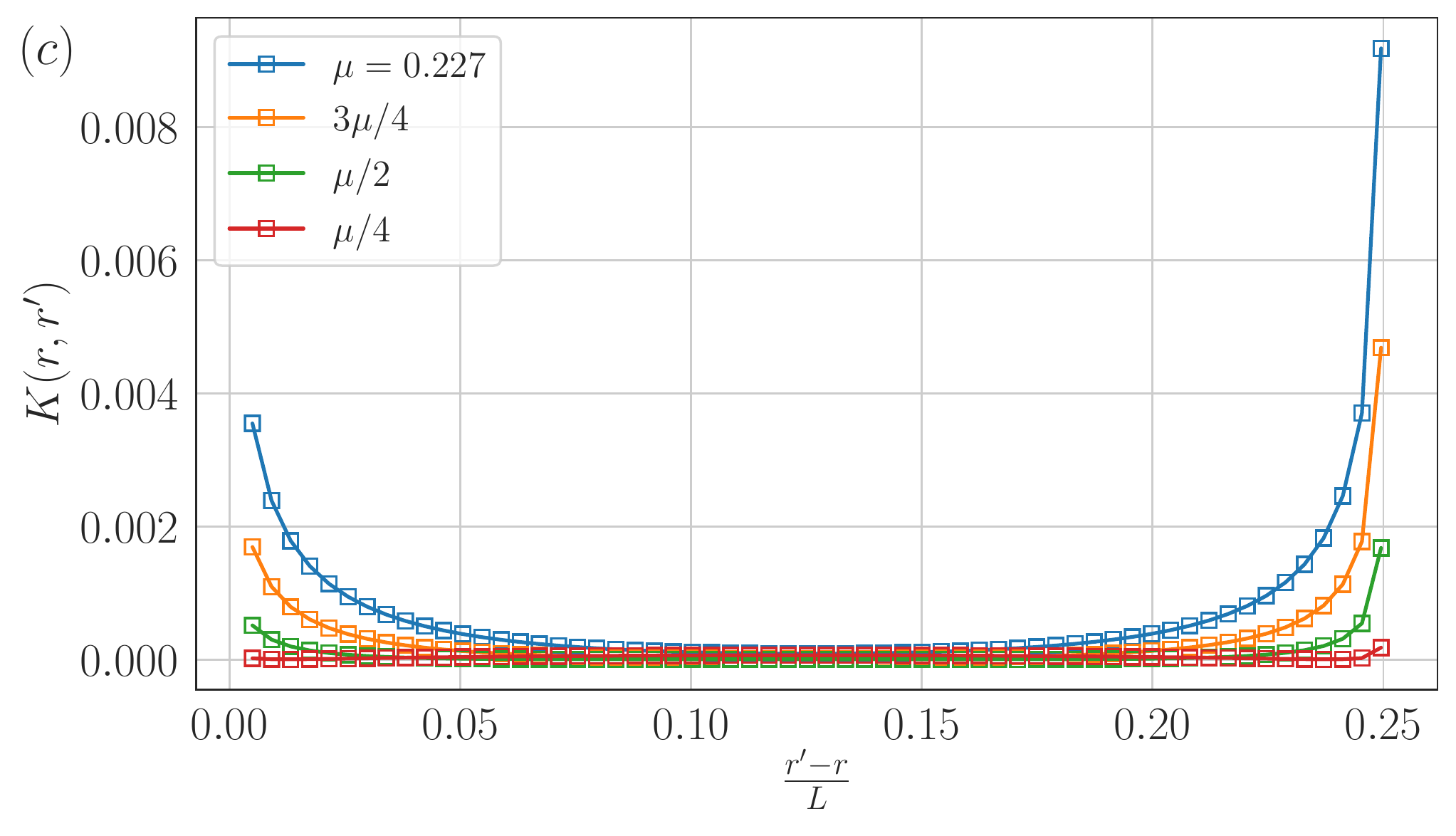}
  \caption{(a) Multiplier PDF $p(w,r,r')$ determined from numerical integration of Eq. (\ref{eq:cauchy_compound}). Dashed lines correspond to the prediction of the fractional Brownian motion model (\ref{eq:cauchy_fbm}). (b) Cumulative PDF of the multipliers $P(w,r,r')$, dashed lines correspond to the fBm prediction $P_{fBm}(w,r,r')=\frac{1}{2}+\frac{1}{\pi}\arctan\left[\frac{w-w_0(r,r')}{\gamma(r,r')} \right]$ with scale and location parameter defined in Eqs. (\ref{eq:scale}) and
  (\ref{eq:location}).
  (c) Kullback-Leibler divergence (\ref{eq:kl}) as a function of the scale separation $r'-r$ for different values of the intermittency coefficient $\mu$. Deviations between the distributions become visible at small and large scale separations and are increasing with increasing intermittency coefficient $\mu$.}
  \label{fig:multiplier}
\end{figure}
We have numerically evaluated the integral in Eq. (\ref{eq:cauchy_compound}) for fixed $r'=0.25L$ and for different values of small-scale $r$. As shown in Fig.~\ref{fig:multiplier}(a), the multiplier PDF of the Gaussian scale mixture is in good agreement with the prediction of the fBm model (dashed lines) given by Eq. (\ref{eq:cauchy_fbm}). At larger scale separations, however, slight differences in the tails can be perceived. These differences manifest themselves in the cumulative distribution function $P(w,r,r')$ in Fig.~\ref{fig:multiplier}(b) as well. In order to further quantify the intermittency corrections,
we have computed the Kullback-Leibler divergence~\cite{kullback1951information}
\begin{equation}
  K(r,r') = \int \textrm{d}w p(w,r,r')\log\left(\frac{p(w,r,r')}{p_{fBm}(w,r,r')} \right)\;,
  \label{eq:kl}
\end{equation}
between the compound Cauchy distribution and the prediction by the fBm model (\ref{eq:cauchy_fbm}) for different intermittency coefficients $\mu$ and $r$ (with fixed large-scale $r'=0.25L$). Although moments of the Cauchy distributions are not defined, it is possible to compute the Kullback-Leibler divergence of two Cauchy distributions~\cite{chyzak2019closed}, which should be sufficient to justify using it for our purpose here.
As expected, the Kullback-Leibler divergence depicted in Fig.~\ref{fig:multiplier}(c) suggests deviations for large scale separations $r'-r$, but decreases for smaller intermittency coefficients $\mu$. For intermediate scale separations, the fBm approximation holds regardless of the magnitude of $\mu$, whereas deviations reappear  for $r \approx r'$, as the Cauchy distribution approaches the Dirac delta-function. Hence, although the multiplier PDF appears to be rather insensitive to intermittency corrections, the multiplier PDF is not fully monofractal as suggested by experimental studies~\cite{chen2003kolmogorov,siefert2007complete}. Due to the Gaussian self-similar features contained in the velocity increments, the multiplier PDF is already so heavy-tailed that intermittency corrections are barely observable. For a proper multiscale description of the turbulent energy cascade, it might thus be appropriate to consider other observables such as transition PDFs~\cite{Friedrich1997} or multipoint moments~\cite{Lvov1996,Eyink1993,friedrich2018multiscale}.

%
%

\section{Extensions and generalizations of the Gaussian scale mixture multipoint statistics}
\label{sec:3d}
In the previous sections we have presented a one-dimensional multipoint statistics of a Gaussian scale mixture which exhibits non-Gaussian features. In the following, we focus on multi-dimensional generalizations as well as further potential improvements.

The $n$-point PDF (\ref{eq:ensemble_n_point}) can readily be generalized to the multipoint statistics of a three-dimensional velocity field according to
\begin{align}\nonumber
 \lefteqn{f_n(\mathbf{u}_1,\mathbf{x}_1;\ldots;\mathbf{u}_n,\mathbf{x}_n)}\\
 =& \int \textrm{d}\xi g(\xi)
 \frac{1}{\sqrt{(2\pi)^3 \det \sigma_\xi}}e^{-\frac{1}{2} u_{i,\alpha} \sigma_{\xi,i,\alpha;j,\beta}^{-1} u_{j,\beta}}\;,
 \label{eq:n_point_3d}
\end{align}
where greek indices indicate spatial components $\alpha=1,2,3$, and where we introduce the covariance matrix $\sigma_{\xi,i,\alpha;j,\beta}=C_{\xi,\alpha \beta}(\mathbf{x}_i-\mathbf{x}_j)$ with the correlation function
\begin{equation}
  C_{\xi,\alpha \beta}(\mathbf{r}) = \left(C_{\xi,rr}(r)-C_{\xi,tt}(r)\right)\frac{r_\alpha r_\beta}{r^2}+C_{\xi,tt}(r)\delta_{\alpha \beta}\;.
  \label{eq:corr_tensor}
\end{equation}
Here, we assume that either the longitudinal $C_{\xi,rr}(r)$ or the transverse correlation function $C_{\xi,tt}(r)$  obey Eq. (\ref{eq:corr_lambda}).
A relation between both functions can then be established by the incompressibility condition $\nabla_\mathbf{x}\cdot \mathbf{u}(\mathbf{x})=0$ for the velocity field. This
leads to the following relation:
\begin{equation}
  \int \textrm{d}\xi g(\xi)\left[ C_{\xi,tt}(r) - \frac{1}{2r}\frac{\partial}{\partial r}  \left(r^2 C_{\xi,rr}(r)\right)  \right]=0\;,
  \label{eq:karman}
\end{equation}
which can be considered as a generalization of the von-K\'arm\'an-Howarth relation~\cite{de1938statistical}. In the context of turbulence,
it remains unclear whether longitudinal and transverse statistical
quantities exhibit different scaling properties~\cite{gotoh-fukayama-etal:2002,shen2002longitudinal,
boratav-pelz:1997,ishihara-gotoh-etal:2009,grauer-homann-pinton:2012,iyer2017reynolds,
Friedrich2016}.
Establishing connections
between longitudinal and transverse statistics on the basis of the integral
constraint (\ref{eq:karman}) will be a task for future work.

Nevertheless, other potential requirements for the statistics such as anisotropies due to the existence of mean fields can be incorporated in the framework by an alternative form of the defining correlation tensor in Eq. (\ref{eq:corr_tensor})~\cite{robertson_1940,1950}. For the case of other turbulent systems such as passive scalars, Rayleigh-B\'enard convection or magnetohydrodynamic (MHD) turbulence, multipoint statistics of different fields can be generated by using cross correlation functions in the Gaussian ansatz, e.g., in the case of MHD, the cross correlation between velocity and magnetic field~\cite{atmos11040382}.
Moreover, spatiotemporal statistics of the velocity field $\mathbf{u}(\mathbf{x},t)$ could be introduced by generalizing the ansatz for the correlation functions to $C_{\xi,ij}(r, \tau)$. Depending on how spatial and temporal fluctuations are distributed, one could thus modify the correlation function in Eq. (\ref{eq:corr_lambda}) accordingly.
Furthermore, due to the underlying fractional Ornstein-Uhlenbeck processes, which dictate the monofractal behavior, the framework should also be general enough to include other phenomenological models of turbulence. The She-Leveque model~\cite{Dubrulle:1994ta}, for instance, could be reproduced by choosing a log-Poisson distribution instead of the lognormal distribution (\ref{eq:lognormal}), possibly in combination with another scale parametrization for $\rho_\xi(r)$ in Eq. (\ref{eq:rho_lambda}).
\section{Outlook}
\label{sec:outlook}
We have proposed an explicit form of the joint multipoint PDF of a multifractal field. The proposed methodology relies on three different elements: \emph{i.)} the stationary solution of a fractional Ornstein-Uhlenbeck process
with Hurst exponent $H$, \emph{ii.)} a re-parametrization $r \rightarrow \rho_{\xi}(r)$ of the scale separation in the corresponding correlation function $C_{\xi}(r)$, and, \emph{iii.)} introduction of fluctuations of the parameter $\xi$.
As highlighted by the joint multipoint PDF (\ref{eq:ensemble_n_point}), these three elements introduce fluctuating correlations which are determined by the parameter or mixing distribution $g(\xi)$.
The proposed Gaussian scale mixture (\ref{eq:ensemble_n_point}) has been used to generate synthetic multifractal fields. Statistical quantities such as the one-point PDF (which is Gaussian due to the underlying fractional Ornstein-Uhlenbeck process), the increment PDF (two-point quantity, which exhibits non-Gaussian features at small scales), the PDF of increment multipliers (three-point quantity), and even the statistics of longitudinal and transverse increments have been reproduced and agree well with empirical findings from turbulence research. Hence, the joint multipoint PDF (\ref{eq:n_point_3d}) defines an appropriate statistical model of a turbulent flow~\cite{monin}.

Future work should address the inclusion of  skewness of the velocity increment distribution into the framework which would thus make it more truthful to the original turbulence problem (i.e., the 4/5-law derived from the Navier-Stokes equation). This can be done by the introduction of a mean in the Gaussian ansatz (\ref{eq:charac_gauss}), which has already been discussed in the context of superstatistics~\cite{sosa2019emergence,NAERT199873}. Moreover, dissipative effects might by incorporated by embedding the Ornstein-Uhlenbeck process which effectively leads to the differentiability of the velocity field~\cite{Sawford_1991,Viggiano2019}.

Due to its straightforward generalizations to multiple dimensions outlined in Sec.~\ref{sec:3d}, the model is directly applicable to the specific tailoring of turbulent wind fields. A central problem in the field of wind energy is the interpolation of sparse wind field measurements~\cite{https://doi.org/10.1002/we.422}. It would thus be desirable to incorporate multiscaling features into a recently developed stochastic interpolation scheme of sparse measurements by fractional Brownian motion~\cite{friedrich2020stochastic}. Moreover, a multiscale refinement method~\cite{Sinhuber_2021}, which is based on the empirical knowledge of the three-point PDF, could be modeled by the compound distribution (\ref{eq:ensemble_n_point}) for $n=3$. Instead of drawing
velocity field realizations similar to the one in Fig.~\ref{fig:u_k62} from the full $n$-point statistics, latter refinement method could thus be used to generate surrogate data at a relatively low computational cost (although deviations from the K62 scaling at small scales might be expected).
On a more basic level, the multipoint statistics (\ref{eq:n_point_3d}) could also be used as a closure method for the multipoint hierarchy derived from the Navier-Stokes equation~\cite{Lundgren1967,friedrich2020non}.
Other potential applications include the modeling of financial data~\cite{ghashghaie1996turbulent}, geophysical settings (non-Gaussian features in the measurements of hydraulic conductivity~\cite{https://doi.org/10.1029/2003GL019320} or subsurface hydrology~\cite{molz1997fractional}), as well as the synthesis of astrophysical magnetic fields and their impact on cosmic particle transport~\cite{reichherzer-etal:2019,giacalone-jokipii:1999,snodin-shukurov-etal:2016}.

\begin{acknowledgements}
J.F. acknowledges funding from the Humboldt Foundation within a Feodor-Lynen fellowship and benefited from financial support through the Project IDEXLYON of the University of Lyon in the framework of the French program ``Programme Investissements d'Avenir'' (ANR-16-IDEX-0005).
\end{acknowledgements}

\begin{appendix}
\section{Derivation of the covariance of the fractional Ornstein-Uhlenbeck process}
\label{app:fOU}
In this appendix we want to derive the stationary limit of the covariance of the fractional Ornstein-Uhlenbeck process (\ref{eq:fOU}), i.e., Eq. (\ref{eq:corr_lambda}) with $\rho_{\xi}(r)$ set to $r$.
The stationary solution of Eq. (\ref{eq:fOU}) reads
\begin{equation}
  w(x) = \int_{-\infty}^x \textrm{d}B^H(x')e^{-(x-x')/L}\;,
\end{equation}
which can be integrated by parts according to
\begin{equation}
  w(x) =  B^H(x) - \frac{1}{L} \int_{-\infty}^x \textrm{d}x' e^{-(x-x')/L} B^H(x') \;.
\end{equation}
The covariance of the fractional Ornstein-Uhlenbeck process can thus be calculated from
\begin{widetext}
\begin{align}\nonumber
  \lefteqn{\left \langle w(x_1)w(x_2) \right \rangle =  \left \langle B^H(x_1) B^H(x_2) \right \rangle }\\ \nonumber
  -& \frac{1}{L} \int_{-\infty}^{x_1} \textrm{d}x_1'e^{-(x_1-x_1')/L} \left \langle B^H(x_1') B^H(x_2) \right \rangle -\frac{1}{L} \int_{-\infty}^{x_2} \textrm{d}x_2'e^{-(x_2-x_2')/L} \left \langle B^H(x_1) B^H(x_2') \right \rangle\\
   +&\frac{1}{L^2}
   \int_{-\infty}^{x_1} \textrm{d}x_1'  \int_{-\infty}^{x_2} \textrm{d}x_2'e^{-(x_1-x_1')/L} e^{-(x_2-x_2')/L} \left \langle B^H(x_1') B^H(x_2') \right \rangle\;.
   \label{eq:cov_app}
\end{align}
\end{widetext}
%
%
%
%
We start with the second integral in Eq. (\ref{eq:cov_app})  and obtain
\begin{widetext}
  \begin{align}\nonumber
    \lefteqn{ \int_{-\infty}^{x_2}  \textrm{d}x_2'e^{x_2'/L} \left(|x_1|^{2H}+|x_2'|^{2H}-|x_1-x_2'|^{2H} \right)} \\ \nonumber
    =& L e^{x_2/L}|x_1|^{2H} + \int_{-\infty}^0 \textrm{d}x_2'e^{x_2'/L} |x_2'|^{2H} + \int_{0}^{x_2} \textrm{d}x_2'e^{x_2'/L} |x_2'|^{2H}
    - \int_{-\infty}^{x_1} \textrm{d}x_2'e^{x_2'/L} (x_1-x_2')^{2H}
    - \int_{x_1}^0 \textrm{d}x_2'e^{x_2'/L} (x_2'-x_1)^{2H} \\ \nonumber
    =& L e^{x_2/L}|x_1|^{2H} + \int_0^{\infty} \textrm{d}x_2'e^{-x_2'/L} (x_2')^{2H} + |x_2|^{2H+1} \int_0^1 \textrm{d}t e^{x_2t/L}t^{2H}
    - \int_{0}^{\infty} \textrm{d}r e^{(x_1-r)/L}r^{2H}-\int_{0}^{x_2-x_1} \textrm{d}r e^{(x_1+r)/L}r^{2H}\\ \nonumber
    =& L e^{x_2/L}|x_1|^{2H} + \Gamma(2H+1)L^{2H+1} + \frac{|x_2|^{2H+1}}{2H+1}
    {_1}F_1 \left(2H+1,2H+2,\frac{x_2}{L}\right) - e^{x_1/L} \Gamma(2H+1)L^{2H+1}\\ \nonumber
    &- e^{x_1/L}|x_2-x_1|^{2H+1} \int_0^1 \textrm{d}t e^{(x_2-x_1)t}t^{2H}\\ \nonumber
    =&  L e^{x_2/L}|x_1|^{2H} + \Gamma(2H+1)L^{2H+1} \left(1- e^{x_1/L}\right) + \frac{|x_2|^{2H+1}}{2H+1}
    {_1}F_1 \left(2H+1,2H+2,\frac{x_2}{L}\right) \\
     &-\frac{|x_2-x_1|^{2H+1}}{2H+1}
    {_1}F_1 \left(2H+1,2H+2,\frac{x_2-x_1}{L}\right)\;.
    \label{eq:int2}
  \end{align}
\end{widetext}
Similarly, the first integral reads
\begin{widetext}
  \begin{align}\nonumber
    \lefteqn{ \int_{-\infty}^{x_1}  \textrm{d}x_1'e^{x_1'/L} \left(|x_1'|^{2H}+|x_2|^{2H}-|x_1'-x_2|^{2H} \right)} \\ \nonumber
    =&  L e^{x_1/L}|x_2|^{2H} + \Gamma(2H+1)L^{2H+1} \left(1- e^{x_2/L}\right) + \frac{|x_1|^{2H+1}}{2H+1}
    {_1}F_1 \left(2H+1,2H+2,\frac{x_1}{L}\right) \\
     &+\frac{|x_2-x_1|^{2H+1}}{2H+1}
    {_1}F_1 \left(2H+1,2H+2,-\frac{x_2-x_1}{L}\right)\;.
    \label{eq:int1}
  \end{align}
\end{widetext}
The double integral in Eq. (\ref{eq:cov_app}) is treated according to
\begin{widetext}
  \begin{align} \nonumber
    \lefteqn{\int_{-\infty}^{x_1} \textrm{d}x_1'  \int_{-\infty}^{x_2} \textrm{d}x_2'
       e^{(x_1'+x_2')/L} \left(|x_1'|^{2H}+|x_2'|^{2H}-|x_1'-x_2'|^{2H} \right) }\\ \nonumber
       =& Le^{x_2/L} \left(\int_{-\infty}^0 \textrm{d}x_1'  e^{x_1'/L}(-x_1')^{2H}+ \int_{0}^{x_1} \textrm{d}x_1'  e^{x_1'/L}(x_1')^{2H}\right)+
       Le^{x_1/L} \left(\int_{-\infty}^0 \textrm{d}x_2'  e^{x_2'/L}(-x_2')^{2H}+ \int_{0}^{x_2} \textrm{d}x_2'  e^{x_2'/L}x_2'^{2H}\right) \\ \nonumber
       & - \int_{-\infty}^{x_1} \textrm{d}x_1'  \int_{-\infty}^{x_1'} \textrm{d}x_2'
          e^{(x_1'+x_2')/L} (x_1'-x_2')^{2H}
          - \int_{-\infty}^{x_1} \textrm{d}x_1'  \int_{x_1'}^{x_2} \textrm{d}x_2'
             e^{(x_1'+x_2')/L} (x_2'-x_1')^{2H} \\ \nonumber
        =& Le^{x_2/L} \left[\Gamma(2H+1) L^{2H+1}+ \frac{|x_1|^{2H+1}}{2H+1}
       {_1}F_1 \left(2H+1,2H+2,\frac{x_1}{L}\right) \right] \\ \nonumber
       &+
       Le^{x_1/L} \left[\Gamma(2H+1) L^{2H+1}+ \frac{|x_2|^{2H+1}}{2H+1}
      {_1}F_1 \left(2H+1,2H+2,\frac{x_2}{L}\right) \right] \\
      &- \underbrace{\int_{-\infty}^{x_1} \textrm{d}x_1' \int_0^{\infty} \textrm{d}r e^{(2x_1'-r)/L}r^{2H}}_{=\frac{L^{2H+2}}{2}\Gamma(2H+1)e^{2x_1/L}} -  \int_{-\infty}^{x_1} \textrm{d}x_1' \int_{0}^{x_2-x_1'} \textrm{d}r e^{(2x_1'+r)/L}r^{2H}\;.
      \label{eq:int3}
  \end{align}
\end{widetext}
Here, the remaining double integral can be evaluated by a partial integration which yields
\begin{widetext}
  \begin{align} \nonumber
    \lefteqn{\int_{-\infty}^{x_1} \textrm{d}x_1' \int_{0}^{x_2-x_1'} \textrm{d}r e^{(2x_1'+r)/L}r^{2H}= \frac{L}{2} \int_{0}^{x_2-x_1} \textrm{d}r e^{(2x_1+r)/L}r^{2H}
    + \frac{L}{2} \int_{-\infty}^{x_1} \textrm{d}x_1' e^{(x_2+x_1')/L} (x_2-x_1')^{2H}} \\ \nonumber
    =& \frac{L}{2}e^{2x_1/L} \frac{|x_2-x_1|^{2H+1}}{2H+1}
   {_1}F_1 \left(2H+1,2H+2,\frac{x_2-x_1}{L}\right)-\frac{L}{2} e^{2x_2/L}\left[\int_{\infty}^0 \textrm{d}r e^{-r/L} r^{2H}
   + \int_0^{x_2-x_1} \textrm{d}r e^{-r/L} r^{2H} \right] \\ \nonumber
   =& \frac{L}{2}e^{2x_1/L} \frac{|x_2-x_1|^{2H+1}}{2H+1}
  {_1}F_1 \left(2H+1,2H+2,\frac{x_2-x_1}{L}\right) \\ \nonumber
  &+\frac{L^{2H+2}}{2} e^{2x_2/L} \Gamma(2H+1)
  -\frac{L}{2}e^{2x_2/L} \frac{|x_2-x_1|^{2H+1}}{2H+1}
 {_1}F_1 \left(2H+1,2H+2,\frac{x_1-x_2}{L}\right)\;.
 \label{eq:int4}
  \end{align}
\end{widetext}
Inserting the terms (\ref{eq:int2})-(\ref{eq:int4}) into the correlation function (\ref{eq:cov_app}) yields
\begin{widetext}
  \begin{align}\nonumber
    \lefteqn{\left \langle w(x_1)w(x_2) \right \rangle
    = -\frac{\sigma^2}{2} (x_2-x_1)^{2H}  +\frac{\sigma^2 L^{2H}}{2}\Gamma(2H+1)\cosh \frac{x_2-x_1}{L}}\\
    &+ \frac{\sigma^2 (x_2-x_1)^{2H+1} }{4L(2H+1)}  \left[
    e^{-\frac{x_2-x_1}{L}}  {_1}F_1 \left(2H+1,2H+2,\frac{x_2-x_1}{L}\right)) -e^{\frac{x_2-x_1}{L}} {_1}F_1 \left(2H+1,2H+2,-\frac{x_2-x_1}{L}\right) \right]\;.
  \end{align}
\end{widetext}
where we assumed that $x_1 \le x_2$.
\end{appendix}
\bibliographystyle{apsrev4-1}
\bibliography{k62_multi.bib}

\begin{thebibliography}{90}%
\makeatletter
\providecommand \@ifxundefined [1]{%
 \@ifx{#1\undefined}
}%
\providecommand \@ifnum [1]{%
 \ifnum #1\expandafter \@firstoftwo
 \else \expandafter \@secondoftwo
 \fi
}%
\providecommand \@ifx [1]{%
 \ifx #1\expandafter \@firstoftwo
 \else \expandafter \@secondoftwo
 \fi
}%
\providecommand \natexlab [1]{#1}%
\providecommand \enquote  [1]{``#1''}%
\providecommand \bibnamefont  [1]{#1}%
\providecommand \bibfnamefont [1]{#1}%
\providecommand \citenamefont [1]{#1}%
\providecommand \href@noop [0]{\@secondoftwo}%
\providecommand \href [0]{\begingroup \@sanitize@url \@href}%
\providecommand \@href[1]{\@@startlink{#1}\@@href}%
\providecommand \@@href[1]{\endgroup#1\@@endlink}%
\providecommand \@sanitize@url [0]{\catcode `\\12\catcode `\$12\catcode
  `\&12\catcode `\#12\catcode `\^12\catcode `\_12\catcode `\%12\relax}%
\providecommand \@@startlink[1]{}%
\providecommand \@@endlink[0]{}%
\providecommand \url  [0]{\begingroup\@sanitize@url \@url }%
\providecommand \@url [1]{\endgroup\@href {#1}{\urlprefix }}%
\providecommand \urlprefix  [0]{URL }%
\providecommand \Eprint [0]{\href }%
\providecommand \doibase [0]{http://dx.doi.org/}%
\providecommand \selectlanguage [0]{\@gobble}%
\providecommand \bibinfo  [0]{\@secondoftwo}%
\providecommand \bibfield  [0]{\@secondoftwo}%
\providecommand \translation [1]{[#1]}%
\providecommand \BibitemOpen [0]{}%
\providecommand \bibitemStop [0]{}%
\providecommand \bibitemNoStop [0]{.\EOS\space}%
\providecommand \EOS [0]{\spacefactor3000\relax}%
\providecommand \BibitemShut  [1]{\csname bibitem#1\endcsname}%
\let\auto@bib@innerbib\@empty
\bibitem [{\citenamefont {Mandelbrot}(1982)}]{mandelbrot1982fractal}%
  \BibitemOpen
  \bibfield  {author} {\bibinfo {author} {\bibfnamefont {B.~B.}\ \bibnamefont
  {Mandelbrot}},\ }\href@noop {} {\emph {\bibinfo {title} {The fractal geometry
  of nature}}},\ Vol.~\bibinfo {volume} {1}\ (\bibinfo  {publisher} {WH Freeman
  New York},\ \bibinfo {year} {1982})\BibitemShut {NoStop}%
\bibitem [{\citenamefont {Barenblatt}(1996)}]{barenblatt1996scaling}%
  \BibitemOpen
  \bibfield  {author} {\bibinfo {author} {\bibfnamefont {G.~I.}\ \bibnamefont
  {Barenblatt}},\ }\href@noop {} {\emph {\bibinfo {title} {Scaling,
  self-similarity, and intermediate asymptotics: dimensional analysis and
  intermediate asymptotics}}},\ \bibinfo {number} {14}\ (\bibinfo  {publisher}
  {Cambridge University Press},\ \bibinfo {year} {1996})\BibitemShut {NoStop}%
\bibitem [{\citenamefont {Barenblatt}\ and\ \citenamefont
  {Zel'Dovich}(1972)}]{barenblatt1972self}%
  \BibitemOpen
  \bibfield  {author} {\bibinfo {author} {\bibfnamefont {G.}~\bibnamefont
  {Barenblatt}}\ and\ \bibinfo {author} {\bibfnamefont {Y.~B.}\ \bibnamefont
  {Zel'Dovich}},\ }\href@noop {} {\bibfield  {journal} {\bibinfo  {journal}
  {Annu. Rev. Fluid Mech.}\ }\textbf {\bibinfo {volume} {4}},\ \bibinfo {pages}
  {285} (\bibinfo {year} {1972})}\BibitemShut {NoStop}%
\bibitem [{\citenamefont {Goldenfeld}(2018)}]{goldenfeld2018lectures}%
  \BibitemOpen
  \bibfield  {author} {\bibinfo {author} {\bibfnamefont {N.}~\bibnamefont
  {Goldenfeld}},\ }\href@noop {} {\emph {\bibinfo {title} {Lectures on phase
  transitions and the renormalization group}}}\ (\bibinfo  {publisher} {CRC
  Press},\ \bibinfo {year} {2018})\BibitemShut {NoStop}%
\bibitem [{\citenamefont {Molz}\ \emph {et~al.}(1997)\citenamefont {Molz},
  \citenamefont {Liu},\ and\ \citenamefont {Szulga}}]{molz1997fractional}%
  \BibitemOpen
  \bibfield  {author} {\bibinfo {author} {\bibfnamefont {F.}~\bibnamefont
  {Molz}}, \bibinfo {author} {\bibfnamefont {H.}~\bibnamefont {Liu}}, \ and\
  \bibinfo {author} {\bibfnamefont {J.}~\bibnamefont {Szulga}},\ }\href@noop {}
  {\bibfield  {journal} {\bibinfo  {journal} {Water. Resour. Res.}\ }\textbf
  {\bibinfo {volume} {33}},\ \bibinfo {pages} {2273} (\bibinfo {year}
  {1997})}\BibitemShut {NoStop}%
\bibitem [{\citenamefont {Molz}\ and\ \citenamefont
  {Boman}(1993)}]{molz1993fractal}%
  \BibitemOpen
  \bibfield  {author} {\bibinfo {author} {\bibfnamefont {F.~J.}\ \bibnamefont
  {Molz}}\ and\ \bibinfo {author} {\bibfnamefont {G.~K.}\ \bibnamefont
  {Boman}},\ }\href@noop {} {\bibfield  {journal} {\bibinfo  {journal} {Water.
  Resour. Res.}\ }\textbf {\bibinfo {volume} {29}},\ \bibinfo {pages} {3769}
  (\bibinfo {year} {1993})}\BibitemShut {NoStop}%
\bibitem [{\citenamefont {Goldstein}\ \emph {et~al.}(1995)\citenamefont
  {Goldstein}, \citenamefont {Roberts},\ and\ \citenamefont
  {Matthaeus}}]{Goldstein1995}%
  \BibitemOpen
  \bibfield  {author} {\bibinfo {author} {\bibfnamefont {M.~L.}\ \bibnamefont
  {Goldstein}}, \bibinfo {author} {\bibfnamefont {D.~A.}\ \bibnamefont
  {Roberts}}, \ and\ \bibinfo {author} {\bibfnamefont {W.}~\bibnamefont
  {Matthaeus}},\ }\href@noop {} {\bibfield  {journal} {\bibinfo  {journal}
  {Annu. Rev. Astron. Astrophys.}\ }\textbf {\bibinfo {volume} {33}},\ \bibinfo
  {pages} {283} (\bibinfo {year} {1995})}\BibitemShut {NoStop}%
\bibitem [{\citenamefont {Makarava}\ \emph {et~al.}(2014)\citenamefont
  {Makarava}, \citenamefont {Menz}, \citenamefont {Theves}, \citenamefont
  {Huisinga}, \citenamefont {Beta},\ and\ \citenamefont
  {Holschneider}}]{Makarava2014}%
  \BibitemOpen
  \bibfield  {author} {\bibinfo {author} {\bibfnamefont {N.}~\bibnamefont
  {Makarava}}, \bibinfo {author} {\bibfnamefont {S.}~\bibnamefont {Menz}},
  \bibinfo {author} {\bibfnamefont {M.}~\bibnamefont {Theves}}, \bibinfo
  {author} {\bibfnamefont {W.}~\bibnamefont {Huisinga}}, \bibinfo {author}
  {\bibfnamefont {C.}~\bibnamefont {Beta}}, \ and\ \bibinfo {author}
  {\bibfnamefont {M.}~\bibnamefont {Holschneider}},\ }\href@noop {} {\bibfield
  {journal} {\bibinfo  {journal} {Phys. Rev. E}\ }\textbf {\bibinfo {volume}
  {90}},\ \bibinfo {pages} {042703} (\bibinfo {year} {2014})}\BibitemShut
  {NoStop}%
\bibitem [{\citenamefont {Peng}\ \emph {et~al.}(1993)\citenamefont {Peng},
  \citenamefont {Mietus}, \citenamefont {Hausdorff}, \citenamefont {Havlin},
  \citenamefont {Stanley},\ and\ \citenamefont {Goldberger}}]{Peng1993}%
  \BibitemOpen
  \bibfield  {author} {\bibinfo {author} {\bibfnamefont {C.-K.}\ \bibnamefont
  {Peng}}, \bibinfo {author} {\bibfnamefont {J.}~\bibnamefont {Mietus}},
  \bibinfo {author} {\bibfnamefont {J.~M.}\ \bibnamefont {Hausdorff}}, \bibinfo
  {author} {\bibfnamefont {S.}~\bibnamefont {Havlin}}, \bibinfo {author}
  {\bibfnamefont {H.~E.}\ \bibnamefont {Stanley}}, \ and\ \bibinfo {author}
  {\bibfnamefont {A.~L.}\ \bibnamefont {Goldberger}},\ }\href@noop {}
  {\bibfield  {journal} {\bibinfo  {journal} {Phys. Rev. Lett.}\ }\textbf
  {\bibinfo {volume} {70}},\ \bibinfo {pages} {1343} (\bibinfo {year}
  {1993})}\BibitemShut {NoStop}%
\bibitem [{\citenamefont {L{\'e}vy}\ and\ \citenamefont
  {Loeve}(1965)}]{levy1965processus}%
  \BibitemOpen
  \bibfield  {author} {\bibinfo {author} {\bibfnamefont {P.}~\bibnamefont
  {L{\'e}vy}}\ and\ \bibinfo {author} {\bibfnamefont {M.}~\bibnamefont
  {Loeve}},\ }\href@noop {} {\emph {\bibinfo {title} {Processus stochastiques
  et mouvement brownien}}}\ (\bibinfo  {publisher} {Gauthier-Villars Paris},\
  \bibinfo {year} {1965})\BibitemShut {NoStop}%
\bibitem [{\citenamefont {Mandelbrot}\ and\ \citenamefont
  {Ness}(1968)}]{mandelbrot1968}%
  \BibitemOpen
  \bibfield  {author} {\bibinfo {author} {\bibfnamefont {B.~B.}\ \bibnamefont
  {Mandelbrot}}\ and\ \bibinfo {author} {\bibfnamefont {J.~W.~V.}\ \bibnamefont
  {Ness}},\ }\href {http://www.jstor.org/stable/2027184} {\bibfield  {journal}
  {\bibinfo  {journal} {SIAM Review}\ }\textbf {\bibinfo {volume} {10}},\
  \bibinfo {pages} {422} (\bibinfo {year} {1968})}\BibitemShut {NoStop}%
\bibitem [{\citenamefont {Grebenkov}\ \emph {et~al.}(2015)\citenamefont
  {Grebenkov}, \citenamefont {Belyaev},\ and\ \citenamefont
  {Jones}}]{grebenkov2015multiscale}%
  \BibitemOpen
  \bibfield  {author} {\bibinfo {author} {\bibfnamefont {D.~S.}\ \bibnamefont
  {Grebenkov}}, \bibinfo {author} {\bibfnamefont {D.}~\bibnamefont {Belyaev}},
  \ and\ \bibinfo {author} {\bibfnamefont {P.~W.}\ \bibnamefont {Jones}},\
  }\href@noop {} {\bibfield  {journal} {\bibinfo  {journal} {Journal of Physics
  A: Mathematical and Theoretical}\ }\textbf {\bibinfo {volume} {49}},\
  \bibinfo {pages} {043001} (\bibinfo {year} {2015})}\BibitemShut {NoStop}%
\bibitem [{\citenamefont {Frisch}(1995)}]{frisch:1995}%
  \BibitemOpen
  \bibfield  {author} {\bibinfo {author} {\bibfnamefont {U.}~\bibnamefont
  {Frisch}},\ }\href@noop {} {\emph {\bibinfo {title} {{Turbulence}}}}\
  (\bibinfo  {publisher} {Cambridge University Press},\ \bibinfo {year}
  {1995})\BibitemShut {NoStop}%
\bibitem [{\citenamefont {Ghashghaie}\ \emph {et~al.}(1996)\citenamefont
  {Ghashghaie}, \citenamefont {Breymann}, \citenamefont {Peinke}, \citenamefont
  {Talkner},\ and\ \citenamefont {Dodge}}]{ghashghaie1996turbulent}%
  \BibitemOpen
  \bibfield  {author} {\bibinfo {author} {\bibfnamefont {S.}~\bibnamefont
  {Ghashghaie}}, \bibinfo {author} {\bibfnamefont {W.}~\bibnamefont
  {Breymann}}, \bibinfo {author} {\bibfnamefont {J.}~\bibnamefont {Peinke}},
  \bibinfo {author} {\bibfnamefont {P.}~\bibnamefont {Talkner}}, \ and\
  \bibinfo {author} {\bibfnamefont {Y.}~\bibnamefont {Dodge}},\ }\href@noop {}
  {\bibfield  {journal} {\bibinfo  {journal} {Nature}\ }\textbf {\bibinfo
  {volume} {381}},\ \bibinfo {pages} {767} (\bibinfo {year}
  {1996})}\BibitemShut {NoStop}%
\bibitem [{\citenamefont {Meerschaert}\ \emph {et~al.}(2004)\citenamefont
  {Meerschaert}, \citenamefont {Kozubowski}, \citenamefont {Molz},\ and\
  \citenamefont {Lu}}]{https://doi.org/10.1029/2003GL019320}%
  \BibitemOpen
  \bibfield  {author} {\bibinfo {author} {\bibfnamefont {M.~M.}\ \bibnamefont
  {Meerschaert}}, \bibinfo {author} {\bibfnamefont {T.~J.}\ \bibnamefont
  {Kozubowski}}, \bibinfo {author} {\bibfnamefont {F.~J.}\ \bibnamefont
  {Molz}}, \ and\ \bibinfo {author} {\bibfnamefont {S.}~\bibnamefont {Lu}},\
  }\href@noop {} {\bibfield  {journal} {\bibinfo  {journal} {Geophys. Res.
  Lett.}\ }\textbf {\bibinfo {volume} {31}} (\bibinfo {year}
  {2004})}\BibitemShut {NoStop}%
\bibitem [{\citenamefont {Hu}\ \emph {et~al.}(2012)\citenamefont {Hu},
  \citenamefont {Cheng}, \citenamefont {Wang},\ and\ \citenamefont
  {Xie}}]{HU2012161}%
  \BibitemOpen
  \bibfield  {author} {\bibinfo {author} {\bibfnamefont {S.}~\bibnamefont
  {Hu}}, \bibinfo {author} {\bibfnamefont {Q.}~\bibnamefont {Cheng}}, \bibinfo
  {author} {\bibfnamefont {L.}~\bibnamefont {Wang}}, \ and\ \bibinfo {author}
  {\bibfnamefont {S.}~\bibnamefont {Xie}},\ }\href {\doibase
  https://doi.org/10.1016/j.apgeog.2011.10.016} {\bibfield  {journal} {\bibinfo
   {journal} {Applied Geography}\ }\textbf {\bibinfo {volume} {34}},\ \bibinfo
  {pages} {161} (\bibinfo {year} {2012})}\BibitemShut {NoStop}%
\bibitem [{\citenamefont
  {Lengyel}(2021)}]{https://doi.org/10.17185/duepublico/73834}%
  \BibitemOpen
  \bibfield  {author} {\bibinfo {author} {\bibfnamefont {J.}~\bibnamefont
  {Lengyel}},\ }\emph {\bibinfo {title} {Multiscale urban analysis and
  modeling: Trends in the Ruhr Area, Germany}},\ \href@noop {} {Ph.D. thesis},\
  \bibinfo  {school} {University of Duisburg-Essen} (\bibinfo {year}
  {2021})\BibitemShut {NoStop}%
\bibitem [{\citenamefont {Juneja}\ \emph {et~al.}(1994)\citenamefont {Juneja},
  \citenamefont {Lathrop}, \citenamefont {Sreenivasan},\ and\ \citenamefont
  {Stolovitzky}}]{juneja1994synthetic}%
  \BibitemOpen
  \bibfield  {author} {\bibinfo {author} {\bibfnamefont {A.}~\bibnamefont
  {Juneja}}, \bibinfo {author} {\bibfnamefont {D.}~\bibnamefont {Lathrop}},
  \bibinfo {author} {\bibfnamefont {K.}~\bibnamefont {Sreenivasan}}, \ and\
  \bibinfo {author} {\bibfnamefont {G.}~\bibnamefont {Stolovitzky}},\
  }\href@noop {} {\bibfield  {journal} {\bibinfo  {journal} {Phys. Rev. E}\
  }\textbf {\bibinfo {volume} {49}},\ \bibinfo {pages} {5179} (\bibinfo {year}
  {1994})}\BibitemShut {NoStop}%
\bibitem [{\citenamefont {Malara}\ \emph {et~al.}(2016)\citenamefont {Malara},
  \citenamefont {Di~Mare}, \citenamefont {Nigro},\ and\ \citenamefont
  {Sorriso-Valvo}}]{malara2016fast}%
  \BibitemOpen
  \bibfield  {author} {\bibinfo {author} {\bibfnamefont {F.}~\bibnamefont
  {Malara}}, \bibinfo {author} {\bibfnamefont {F.}~\bibnamefont {Di~Mare}},
  \bibinfo {author} {\bibfnamefont {G.}~\bibnamefont {Nigro}}, \ and\ \bibinfo
  {author} {\bibfnamefont {L.}~\bibnamefont {Sorriso-Valvo}},\ }\href@noop {}
  {\bibfield  {journal} {\bibinfo  {journal} {Phys. Rev. E}\ }\textbf {\bibinfo
  {volume} {94}},\ \bibinfo {pages} {053109} (\bibinfo {year}
  {2016})}\BibitemShut {NoStop}%
\bibitem [{\citenamefont {Rosales}\ and\ \citenamefont
  {Meneveau}(2008)}]{rosales2008anomalous}%
  \BibitemOpen
  \bibfield  {author} {\bibinfo {author} {\bibfnamefont {C.}~\bibnamefont
  {Rosales}}\ and\ \bibinfo {author} {\bibfnamefont {C.}~\bibnamefont
  {Meneveau}},\ }\href@noop {} {\bibfield  {journal} {\bibinfo  {journal}
  {Phys. Rev. E}\ }\textbf {\bibinfo {volume} {78}},\ \bibinfo {pages} {016313}
  (\bibinfo {year} {2008})}\BibitemShut {NoStop}%
\bibitem [{\citenamefont {Schertzer}\ and\ \citenamefont
  {Lovejoy}(1988)}]{schertzer1988multifractal}%
  \BibitemOpen
  \bibfield  {author} {\bibinfo {author} {\bibfnamefont {D.}~\bibnamefont
  {Schertzer}}\ and\ \bibinfo {author} {\bibfnamefont {S.}~\bibnamefont
  {Lovejoy}},\ }\href@noop {} {\bibfield  {journal} {\bibinfo  {journal}
  {Atmospheric research}\ }\textbf {\bibinfo {volume} {21}},\ \bibinfo {pages}
  {337} (\bibinfo {year} {1988})}\BibitemShut {NoStop}%
\bibitem [{\citenamefont {Bacry}\ \emph {et~al.}(2001)\citenamefont {Bacry},
  \citenamefont {Delour},\ and\ \citenamefont {Muzy}}]{bacry2001multifractal}%
  \BibitemOpen
  \bibfield  {author} {\bibinfo {author} {\bibfnamefont {E.}~\bibnamefont
  {Bacry}}, \bibinfo {author} {\bibfnamefont {J.}~\bibnamefont {Delour}}, \
  and\ \bibinfo {author} {\bibfnamefont {J.-F.}\ \bibnamefont {Muzy}},\
  }\href@noop {} {\bibfield  {journal} {\bibinfo  {journal} {Phys. Rev. E}\
  }\textbf {\bibinfo {volume} {64}},\ \bibinfo {pages} {026103} (\bibinfo
  {year} {2001})}\BibitemShut {NoStop}%
\bibitem [{\citenamefont {Chevillard}\ \emph {et~al.}(2010)\citenamefont
  {Chevillard}, \citenamefont {Robert},\ and\ \citenamefont
  {Vargas}}]{chevillard2010stochastic}%
  \BibitemOpen
  \bibfield  {author} {\bibinfo {author} {\bibfnamefont {L.}~\bibnamefont
  {Chevillard}}, \bibinfo {author} {\bibfnamefont {R.}~\bibnamefont {Robert}},
  \ and\ \bibinfo {author} {\bibfnamefont {V.}~\bibnamefont {Vargas}},\
  }\href@noop {} {\bibfield  {journal} {\bibinfo  {journal} {EPL}\ }\textbf
  {\bibinfo {volume} {89}},\ \bibinfo {pages} {54002} (\bibinfo {year}
  {2010})}\BibitemShut {NoStop}%
\bibitem [{\citenamefont {Chevillard}\ \emph {et~al.}(2019)\citenamefont
  {Chevillard}, \citenamefont {Garban}, \citenamefont {Rhodes},\ and\
  \citenamefont {Vargas}}]{Chevillard_2019}%
  \BibitemOpen
  \bibfield  {author} {\bibinfo {author} {\bibfnamefont {L.}~\bibnamefont
  {Chevillard}}, \bibinfo {author} {\bibfnamefont {C.}~\bibnamefont {Garban}},
  \bibinfo {author} {\bibfnamefont {R.}~\bibnamefont {Rhodes}}, \ and\ \bibinfo
  {author} {\bibfnamefont {V.}~\bibnamefont {Vargas}},\ }\href {\doibase
  10.1007/s00023-019-00842-y} {\bibfield  {journal} {\bibinfo  {journal} {Ann.
  Henri Poincar{\'e}}\ }\textbf {\bibinfo {volume} {20}},\ \bibinfo {pages}
  {3693} (\bibinfo {year} {2019})}\BibitemShut {NoStop}%
\bibitem [{\citenamefont {Chevillard}\ \emph {et~al.}(2020)\citenamefont
  {Chevillard}, \citenamefont {Lagoin},\ and\ \citenamefont
  {Roux}}]{chevillard2020multifractal}%
  \BibitemOpen
  \bibfield  {author} {\bibinfo {author} {\bibfnamefont {L.}~\bibnamefont
  {Chevillard}}, \bibinfo {author} {\bibfnamefont {M.}~\bibnamefont {Lagoin}},
  \ and\ \bibinfo {author} {\bibfnamefont {S.~G.}\ \bibnamefont {Roux}},\
  }\href@noop {} {\bibfield  {journal} {\bibinfo  {journal} {arXiv preprint
  arXiv:2011.09503}\ } (\bibinfo {year} {2020})}\BibitemShut {NoStop}%
\bibitem [{\citenamefont {Apolin{\'a}rio}\ and\ \citenamefont
  {Moriconi}(2020)}]{apolinario2020shot}%
  \BibitemOpen
  \bibfield  {author} {\bibinfo {author} {\bibfnamefont {G.~B.}\ \bibnamefont
  {Apolin{\'a}rio}}\ and\ \bibinfo {author} {\bibfnamefont {L.}~\bibnamefont
  {Moriconi}},\ }\href@noop {} {\bibfield  {journal} {\bibinfo  {journal}
  {Journal of Statistical Mechanics: Theory and Experiment}\ }\textbf {\bibinfo
  {volume} {2020}},\ \bibinfo {pages} {073208} (\bibinfo {year}
  {2020})}\BibitemShut {NoStop}%
\bibitem [{\citenamefont {Meneveau}\ \emph {et~al.}(1990)\citenamefont
  {Meneveau}, \citenamefont {Sreenivasan}, \citenamefont {Kailasnath},\ and\
  \citenamefont {Fan}}]{meneveau1990joint}%
  \BibitemOpen
  \bibfield  {author} {\bibinfo {author} {\bibfnamefont {C.}~\bibnamefont
  {Meneveau}}, \bibinfo {author} {\bibfnamefont {K.}~\bibnamefont
  {Sreenivasan}}, \bibinfo {author} {\bibfnamefont {P.}~\bibnamefont
  {Kailasnath}}, \ and\ \bibinfo {author} {\bibfnamefont {M.}~\bibnamefont
  {Fan}},\ }\href@noop {} {\bibfield  {journal} {\bibinfo  {journal} {Phys.
  Rev. A}\ }\textbf {\bibinfo {volume} {41}},\ \bibinfo {pages} {894} (\bibinfo
  {year} {1990})}\BibitemShut {NoStop}%
\bibitem [{\citenamefont {Jaffard}\ \emph {et~al.}(2019)\citenamefont
  {Jaffard}, \citenamefont {Seuret}, \citenamefont {Wendt}, \citenamefont
  {Leonarduzzi},\ and\ \citenamefont {Abry}}]{Jaffard_2019}%
  \BibitemOpen
  \bibfield  {author} {\bibinfo {author} {\bibfnamefont {S.}~\bibnamefont
  {Jaffard}}, \bibinfo {author} {\bibfnamefont {S.}~\bibnamefont {Seuret}},
  \bibinfo {author} {\bibfnamefont {H.}~\bibnamefont {Wendt}}, \bibinfo
  {author} {\bibfnamefont {R.}~\bibnamefont {Leonarduzzi}}, \ and\ \bibinfo
  {author} {\bibfnamefont {P.}~\bibnamefont {Abry}},\ }\href {\doibase
  10.1098/rspa.2019.0150} {\bibfield  {journal} {\bibinfo  {journal} {Proc. R.
  Soc. A}\ }\textbf {\bibinfo {volume} {475}},\ \bibinfo {pages} {20190150}
  (\bibinfo {year} {2019})}\BibitemShut {NoStop}%
\bibitem [{\citenamefont {Monin}\ and\ \citenamefont {Yaglom}(2007)}]{monin}%
  \BibitemOpen
  \bibfield  {author} {\bibinfo {author} {\bibfnamefont {A.~S.}\ \bibnamefont
  {Monin}}\ and\ \bibinfo {author} {\bibfnamefont {A.~M.}\ \bibnamefont
  {Yaglom}},\ }\href@noop {} {\emph {\bibinfo {title} {{Statistical Fluid
  Mechanics: Mechanics of Turbulence}}}}\ (\bibinfo  {publisher} {Courier Dover
  Publications},\ \bibinfo {year} {2007})\BibitemShut {NoStop}%
\bibitem [{\citenamefont {Nawroth}\ and\ \citenamefont
  {Peinke}(2006)}]{nawroth-peinke:2006}%
  \BibitemOpen
  \bibfield  {author} {\bibinfo {author} {\bibfnamefont {A.}~\bibnamefont
  {Nawroth}}\ and\ \bibinfo {author} {\bibfnamefont {J.}~\bibnamefont
  {Peinke}},\ }\href {\doibase
  http://dx.doi.org/10.1016/j.physleta.2006.08.024} {\bibfield  {journal}
  {\bibinfo  {journal} {PLA}\ }\textbf {\bibinfo {volume} {360}},\ \bibinfo
  {pages} {234 } (\bibinfo {year} {2006})}\BibitemShut {NoStop}%
\bibitem [{\citenamefont {Friedrich}\ and\ \citenamefont
  {Peinke}(1997)}]{Friedrich1997}%
  \BibitemOpen
  \bibfield  {author} {\bibinfo {author} {\bibfnamefont {R.}~\bibnamefont
  {Friedrich}}\ and\ \bibinfo {author} {\bibfnamefont {J.}~\bibnamefont
  {Peinke}},\ }\href@noop {} {\bibfield  {journal} {\bibinfo  {journal} {Phys.
  Rev. Lett.}\ }\textbf {\bibinfo {volume} {78}},\ \bibinfo {pages} {863}
  (\bibinfo {year} {1997})}\BibitemShut {NoStop}%
\bibitem [{\citenamefont {Friedrich}\ \emph {et~al.}(2011)\citenamefont
  {Friedrich}, \citenamefont {Peinke}, \citenamefont {Sahimi},\ and\
  \citenamefont {Tabar}}]{Friedrich2011a}%
  \BibitemOpen
  \bibfield  {author} {\bibinfo {author} {\bibfnamefont {R.}~\bibnamefont
  {Friedrich}}, \bibinfo {author} {\bibfnamefont {J.}~\bibnamefont {Peinke}},
  \bibinfo {author} {\bibfnamefont {M.}~\bibnamefont {Sahimi}}, \ and\ \bibinfo
  {author} {\bibfnamefont {R.~M.}\ \bibnamefont {Tabar}},\ }\href {\doibase
  10.1016/j.physrep.2011.05.003} {\bibfield  {journal} {\bibinfo  {journal}
  {Phys. Rep.}\ }\textbf {\bibinfo {volume} {506}},\ \bibinfo {pages} {87}
  (\bibinfo {year} {2011})}\BibitemShut {NoStop}%
\bibitem [{\citenamefont {Friedrich}(2017)}]{Friedrich2017}%
  \BibitemOpen
  \bibfield  {author} {\bibinfo {author} {\bibfnamefont {J.}~\bibnamefont
  {Friedrich}},\ }\emph {\bibinfo {title} {Closure of the
  Lundgren-Monin-Novikov hierarchy in turbulence via a Markov property of
  velocity increments in scale}},\ \href@noop {} {Ph.D. thesis},\ \bibinfo
  {school} {Ruhr-University Bochum} (\bibinfo {year} {2017})\BibitemShut
  {NoStop}%
\bibitem [{\citenamefont {Sinhuber}\ \emph {et~al.}(2021)\citenamefont
  {Sinhuber}, \citenamefont {Friedrich}, \citenamefont {Grauer},\ and\
  \citenamefont {Wilczek}}]{Sinhuber_2021}%
  \BibitemOpen
  \bibfield  {author} {\bibinfo {author} {\bibfnamefont {M.}~\bibnamefont
  {Sinhuber}}, \bibinfo {author} {\bibfnamefont {J.}~\bibnamefont {Friedrich}},
  \bibinfo {author} {\bibfnamefont {R.}~\bibnamefont {Grauer}}, \ and\ \bibinfo
  {author} {\bibfnamefont {M.}~\bibnamefont {Wilczek}},\ }\href@noop {}
  {\bibfield  {journal} {\bibinfo  {journal} {New J. Phys.}\ } (\bibinfo {year}
  {2021})}\BibitemShut {NoStop}%
\bibitem [{\citenamefont {{Di Francesco}}\ and\ \citenamefont
  {Kutasov}(1991)}]{DIFRANCESCO1991385}%
  \BibitemOpen
  \bibfield  {author} {\bibinfo {author} {\bibfnamefont {P.}~\bibnamefont {{Di
  Francesco}}}\ and\ \bibinfo {author} {\bibfnamefont {D.}~\bibnamefont
  {Kutasov}},\ }\href {\doibase https://doi.org/10.1016/0370-2693(91)90444-U}
  {\bibfield  {journal} {\bibinfo  {journal} {Phys. Lett. B}\ }\textbf
  {\bibinfo {volume} {261}},\ \bibinfo {pages} {385} (\bibinfo {year}
  {1991})}\BibitemShut {NoStop}%
\bibitem [{\citenamefont {Babujian}\ \emph {et~al.}(2017)\citenamefont
  {Babujian}, \citenamefont {Karowski},\ and\ \citenamefont
  {Tsvelik}}]{babujian2017multipoint}%
  \BibitemOpen
  \bibfield  {author} {\bibinfo {author} {\bibfnamefont {H.}~\bibnamefont
  {Babujian}}, \bibinfo {author} {\bibfnamefont {M.}~\bibnamefont {Karowski}},
  \ and\ \bibinfo {author} {\bibfnamefont {A.}~\bibnamefont {Tsvelik}},\
  }\href@noop {} {\bibfield  {journal} {\bibinfo  {journal} {Nucl. Phys. B}\
  }\textbf {\bibinfo {volume} {917}},\ \bibinfo {pages} {122} (\bibinfo {year}
  {2017})}\BibitemShut {NoStop}%
\bibitem [{\citenamefont {Kugler}\ \emph {et~al.}(2021)\citenamefont {Kugler},
  \citenamefont {Lee},\ and\ \citenamefont {von Delft}}]{kugler2021multipoint}%
  \BibitemOpen
  \bibfield  {author} {\bibinfo {author} {\bibfnamefont {F.~B.}\ \bibnamefont
  {Kugler}}, \bibinfo {author} {\bibfnamefont {S.-S.~B.}\ \bibnamefont {Lee}},
  \ and\ \bibinfo {author} {\bibfnamefont {J.}~\bibnamefont {von Delft}},\
  }\href@noop {} {\bibfield  {journal} {\bibinfo  {journal} {arXiv preprint
  arXiv:2101.00707}\ } (\bibinfo {year} {2021})}\BibitemShut {NoStop}%
\bibitem [{\citenamefont {Squarcini}(2021)}]{squarcini2021multipoint}%
  \BibitemOpen
  \bibfield  {author} {\bibinfo {author} {\bibfnamefont {A.}~\bibnamefont
  {Squarcini}},\ }\href@noop {} {\bibfield  {journal} {\bibinfo  {journal}
  {arXiv preprint arXiv:2104.05073}\ } (\bibinfo {year} {2021})}\BibitemShut
  {NoStop}%
\bibitem [{\citenamefont {Kitanine}\ \emph {et~al.}(1999)\citenamefont
  {Kitanine}, \citenamefont {Maillet},\ and\ \citenamefont
  {Terras}}]{kitanine1999form}%
  \BibitemOpen
  \bibfield  {author} {\bibinfo {author} {\bibfnamefont {N.}~\bibnamefont
  {Kitanine}}, \bibinfo {author} {\bibfnamefont {J.}~\bibnamefont {Maillet}}, \
  and\ \bibinfo {author} {\bibfnamefont {V.}~\bibnamefont {Terras}},\
  }\href@noop {} {\bibfield  {journal} {\bibinfo  {journal} {Nucl. Phys. B}\
  }\textbf {\bibinfo {volume} {554}},\ \bibinfo {pages} {647} (\bibinfo {year}
  {1999})}\BibitemShut {NoStop}%
\bibitem [{\citenamefont {Lundgren}(1967)}]{Lundgren1967}%
  \BibitemOpen
  \bibfield  {author} {\bibinfo {author} {\bibfnamefont {T.~S.}\ \bibnamefont
  {Lundgren}},\ }\href@noop {} {\bibfield  {journal} {\bibinfo  {journal}
  {Phys. Fluids}\ }\textbf {\bibinfo {volume} {10}},\ \bibinfo {pages} {969}
  (\bibinfo {year} {1967})}\BibitemShut {NoStop}%
\bibitem [{\citenamefont {Mydlarski}\ \emph {et~al.}(1998)\citenamefont
  {Mydlarski}, \citenamefont {Pumir}, \citenamefont {Shraiman}, \citenamefont
  {Siggia},\ and\ \citenamefont {Warhaft}}]{mydlarski1998structures}%
  \BibitemOpen
  \bibfield  {author} {\bibinfo {author} {\bibfnamefont {L.}~\bibnamefont
  {Mydlarski}}, \bibinfo {author} {\bibfnamefont {A.}~\bibnamefont {Pumir}},
  \bibinfo {author} {\bibfnamefont {B.~I.}\ \bibnamefont {Shraiman}}, \bibinfo
  {author} {\bibfnamefont {E.~D.}\ \bibnamefont {Siggia}}, \ and\ \bibinfo
  {author} {\bibfnamefont {Z.}~\bibnamefont {Warhaft}},\ }\href@noop {}
  {\bibfield  {journal} {\bibinfo  {journal} {Phys. Rev. Lett.}\ }\textbf
  {\bibinfo {volume} {81}},\ \bibinfo {pages} {4373} (\bibinfo {year}
  {1998})}\BibitemShut {NoStop}%
\bibitem [{\citenamefont {Yang}\ \emph {et~al.}(2020)\citenamefont {Yang},
  \citenamefont {Pumir},\ and\ \citenamefont {Xu}}]{yang2020dynamics}%
  \BibitemOpen
  \bibfield  {author} {\bibinfo {author} {\bibfnamefont {P.-F.}\ \bibnamefont
  {Yang}}, \bibinfo {author} {\bibfnamefont {A.}~\bibnamefont {Pumir}}, \ and\
  \bibinfo {author} {\bibfnamefont {H.}~\bibnamefont {Xu}},\ }\href@noop {}
  {\bibfield  {journal} {\bibinfo  {journal} {J. Fluid Mech.}\ }\textbf
  {\bibinfo {volume} {897}},\ \bibinfo {pages} {A9} (\bibinfo {year}
  {2020})}\BibitemShut {NoStop}%
\bibitem [{\citenamefont {Stresing}\ and\ \citenamefont
  {Peinke}(2010)}]{Stresing_2010}%
  \BibitemOpen
  \bibfield  {author} {\bibinfo {author} {\bibfnamefont {R.}~\bibnamefont
  {Stresing}}\ and\ \bibinfo {author} {\bibfnamefont {J.}~\bibnamefont
  {Peinke}},\ }\href {\doibase 10.1088/1367-2630/12/10/103046} {\bibfield
  {journal} {\bibinfo  {journal} {New J. Phys.}\ }\textbf {\bibinfo {volume}
  {12}},\ \bibinfo {pages} {103046} (\bibinfo {year} {2010})}\BibitemShut
  {NoStop}%
\bibitem [{\citenamefont {Kolmogorov}(1962)}]{kolmogorov:1962}%
  \BibitemOpen
  \bibfield  {author} {\bibinfo {author} {\bibfnamefont {A.~N.}\ \bibnamefont
  {Kolmogorov}},\ }\href {\doibase 10.1017/S0022112062000518} {\bibfield
  {journal} {\bibinfo  {journal} {J.~Fluid Mech}\ }\textbf {\bibinfo {volume}
  {13}},\ \bibinfo {pages} {82} (\bibinfo {year} {1962})}\BibitemShut {NoStop}%
\bibitem [{\citenamefont {Obukhov}(1962)}]{obukhov:1962}%
  \BibitemOpen
  \bibfield  {author} {\bibinfo {author} {\bibfnamefont {A.}~\bibnamefont
  {Obukhov}},\ }\href@noop {} {\bibfield  {journal} {\bibinfo  {journal} {J.
  Fluid Mech.}\ }\textbf {\bibinfo {volume} {13}},\ \bibinfo {pages} {77}
  (\bibinfo {year} {1962})}\BibitemShut {NoStop}%
\bibitem [{\citenamefont {Beck}(2004)}]{BECK2004195}%
  \BibitemOpen
  \bibfield  {author} {\bibinfo {author} {\bibfnamefont {C.}~\bibnamefont
  {Beck}},\ }\href {\doibase https://doi.org/10.1016/j.physd.2004.01.020}
  {\bibfield  {journal} {\bibinfo  {journal} {Physica D}\ }\textbf {\bibinfo
  {volume} {193}},\ \bibinfo {pages} {195} (\bibinfo {year}
  {2004})}\BibitemShut {NoStop}%
\bibitem [{\citenamefont {Castaing}\ \emph {et~al.}(1990)\citenamefont
  {Castaing}, \citenamefont {Gagne},\ and\ \citenamefont
  {Hopfinger}}]{CASTAING1990177}%
  \BibitemOpen
  \bibfield  {author} {\bibinfo {author} {\bibfnamefont {B.}~\bibnamefont
  {Castaing}}, \bibinfo {author} {\bibfnamefont {Y.}~\bibnamefont {Gagne}}, \
  and\ \bibinfo {author} {\bibfnamefont {E.}~\bibnamefont {Hopfinger}},\ }\href
  {\doibase https://doi.org/10.1016/0167-2789(90)90035-N} {\bibfield  {journal}
  {\bibinfo  {journal} {Physica D}\ }\textbf {\bibinfo {volume} {46}},\
  \bibinfo {pages} {177} (\bibinfo {year} {1990})}\BibitemShut {NoStop}%
\bibitem [{\citenamefont {Chabaud}\ \emph {et~al.}(1994)\citenamefont
  {Chabaud}, \citenamefont {Naert}, \citenamefont {Peinke}, \citenamefont
  {Chilla}, \citenamefont {Castaing},\ and\ \citenamefont
  {Hebral}}]{chabaud1994transition}%
  \BibitemOpen
  \bibfield  {author} {\bibinfo {author} {\bibfnamefont {B.}~\bibnamefont
  {Chabaud}}, \bibinfo {author} {\bibfnamefont {A.}~\bibnamefont {Naert}},
  \bibinfo {author} {\bibfnamefont {J.}~\bibnamefont {Peinke}}, \bibinfo
  {author} {\bibfnamefont {F.}~\bibnamefont {Chilla}}, \bibinfo {author}
  {\bibfnamefont {B.}~\bibnamefont {Castaing}}, \ and\ \bibinfo {author}
  {\bibfnamefont {B.}~\bibnamefont {Hebral}},\ }\href@noop {} {\bibfield
  {journal} {\bibinfo  {journal} {Phys. Rev. Lett.}\ }\textbf {\bibinfo
  {volume} {73}},\ \bibinfo {pages} {3227} (\bibinfo {year}
  {1994})}\BibitemShut {NoStop}%
\bibitem [{\citenamefont {Naert}\ \emph {et~al.}(1998)\citenamefont {Naert},
  \citenamefont {Castaing}, \citenamefont {Chabaud}, \citenamefont
  {H{\'e}bral},\ and\ \citenamefont {Peinke}}]{NAERT199873}%
  \BibitemOpen
  \bibfield  {author} {\bibinfo {author} {\bibfnamefont {A.}~\bibnamefont
  {Naert}}, \bibinfo {author} {\bibfnamefont {B.}~\bibnamefont {Castaing}},
  \bibinfo {author} {\bibfnamefont {B.}~\bibnamefont {Chabaud}}, \bibinfo
  {author} {\bibfnamefont {B.}~\bibnamefont {H{\'e}bral}}, \ and\ \bibinfo
  {author} {\bibfnamefont {J.}~\bibnamefont {Peinke}},\ }\href {\doibase
  https://doi.org/10.1016/S0167-2789(97)00196-6} {\bibfield  {journal}
  {\bibinfo  {journal} {Physica D}\ }\textbf {\bibinfo {volume} {113}},\
  \bibinfo {pages} {73} (\bibinfo {year} {1998})}\BibitemShut {NoStop}%
\bibitem [{\citenamefont {Yakhot}(2006)}]{YAKHOT2006166}%
  \BibitemOpen
  \bibfield  {author} {\bibinfo {author} {\bibfnamefont {V.}~\bibnamefont
  {Yakhot}},\ }\href {\doibase https://doi.org/10.1016/j.physd.2006.01.012}
  {\bibfield  {journal} {\bibinfo  {journal} {Physica D}\ }\textbf {\bibinfo
  {volume} {215}},\ \bibinfo {pages} {166} (\bibinfo {year}
  {2006})}\BibitemShut {NoStop}%
\bibitem [{\citenamefont {Wang}\ \emph {et~al.}(2012)\citenamefont {Wang},
  \citenamefont {Kuo}, \citenamefont {Bae},\ and\ \citenamefont
  {Granick}}]{Wang_2012}%
  \BibitemOpen
  \bibfield  {author} {\bibinfo {author} {\bibfnamefont {B.}~\bibnamefont
  {Wang}}, \bibinfo {author} {\bibfnamefont {J.}~\bibnamefont {Kuo}}, \bibinfo
  {author} {\bibfnamefont {S.~C.}\ \bibnamefont {Bae}}, \ and\ \bibinfo
  {author} {\bibfnamefont {S.}~\bibnamefont {Granick}},\ }\href {\doibase
  10.1038/nmat3308} {\bibfield  {journal} {\bibinfo  {journal} {Nature
  Materials}\ }\textbf {\bibinfo {volume} {11}},\ \bibinfo {pages} {481}
  (\bibinfo {year} {2012})}\BibitemShut {NoStop}%
\bibitem [{\citenamefont {{Leon Chen}}\ and\ \citenamefont
  {Beck}(2008)}]{LEONCHEN20083162}%
  \BibitemOpen
  \bibfield  {author} {\bibinfo {author} {\bibfnamefont {L.}~\bibnamefont
  {{Leon Chen}}}\ and\ \bibinfo {author} {\bibfnamefont {C.}~\bibnamefont
  {Beck}},\ }\href {\doibase https://doi.org/10.1016/j.physa.2008.01.116}
  {\bibfield  {journal} {\bibinfo  {journal} {Physica A}\ }\textbf {\bibinfo
  {volume} {387}},\ \bibinfo {pages} {3162} (\bibinfo {year}
  {2008})}\BibitemShut {NoStop}%
\bibitem [{\citenamefont {Ausloos}\ and\ \citenamefont
  {Ivanova}(2003)}]{ausloos2003dynamical}%
  \BibitemOpen
  \bibfield  {author} {\bibinfo {author} {\bibfnamefont {M.}~\bibnamefont
  {Ausloos}}\ and\ \bibinfo {author} {\bibfnamefont {K.}~\bibnamefont
  {Ivanova}},\ }\href@noop {} {\bibfield  {journal} {\bibinfo  {journal} {Phys.
  Rev. E}\ }\textbf {\bibinfo {volume} {68}},\ \bibinfo {pages} {046122}
  (\bibinfo {year} {2003})}\BibitemShut {NoStop}%
\bibitem [{\citenamefont {Metzler}(2020)}]{metzler2020superstatistics}%
  \BibitemOpen
  \bibfield  {author} {\bibinfo {author} {\bibfnamefont {R.}~\bibnamefont
  {Metzler}},\ }\href@noop {} {\bibfield  {journal} {\bibinfo  {journal} {EPJ}\
  }\textbf {\bibinfo {volume} {229}},\ \bibinfo {pages} {711} (\bibinfo {year}
  {2020})}\BibitemShut {NoStop}%
\bibitem [{\citenamefont {M{\"u}cke}\ \emph {et~al.}(2011)\citenamefont
  {M{\"u}cke}, \citenamefont {Kleinhans},\ and\ \citenamefont
  {Peinke}}]{https://doi.org/10.1002/we.422}%
  \BibitemOpen
  \bibfield  {author} {\bibinfo {author} {\bibfnamefont {T.}~\bibnamefont
  {M{\"u}cke}}, \bibinfo {author} {\bibfnamefont {D.}~\bibnamefont
  {Kleinhans}}, \ and\ \bibinfo {author} {\bibfnamefont {J.}~\bibnamefont
  {Peinke}},\ }\href@noop {} {\bibfield  {journal} {\bibinfo  {journal} {Wind
  Energy}\ }\textbf {\bibinfo {volume} {14}},\ \bibinfo {pages} {301} (\bibinfo
  {year} {2011})}\BibitemShut {NoStop}%
\bibitem [{\citenamefont {Chen}\ \emph {et~al.}(2003)\citenamefont {Chen},
  \citenamefont {Chen}, \citenamefont {Eyink},\ and\ \citenamefont
  {Sreenivasan}}]{chen2003kolmogorov}%
  \BibitemOpen
  \bibfield  {author} {\bibinfo {author} {\bibfnamefont {Q.}~\bibnamefont
  {Chen}}, \bibinfo {author} {\bibfnamefont {S.}~\bibnamefont {Chen}}, \bibinfo
  {author} {\bibfnamefont {G.~L.}\ \bibnamefont {Eyink}}, \ and\ \bibinfo
  {author} {\bibfnamefont {K.~R.}\ \bibnamefont {Sreenivasan}},\ }\href@noop {}
  {\bibfield  {journal} {\bibinfo  {journal} {Phys. Rev. Lett.}\ }\textbf
  {\bibinfo {volume} {90}},\ \bibinfo {pages} {254501} (\bibinfo {year}
  {2003})}\BibitemShut {NoStop}%
\bibitem [{\citenamefont {Beck}\ and\ \citenamefont
  {Cohen}(2003)}]{BECK2003267}%
  \BibitemOpen
  \bibfield  {author} {\bibinfo {author} {\bibfnamefont {C.}~\bibnamefont
  {Beck}}\ and\ \bibinfo {author} {\bibfnamefont {E.}~\bibnamefont {Cohen}},\
  }\href {\doibase https://doi.org/10.1016/S0378-4371(03)00019-0} {\bibfield
  {journal} {\bibinfo  {journal} {Physica A}\ }\textbf {\bibinfo {volume}
  {322}},\ \bibinfo {pages} {267} (\bibinfo {year} {2003})}\BibitemShut
  {NoStop}%
\bibitem [{\citenamefont {Hopf}(1952)}]{Hopf1952}%
  \BibitemOpen
  \bibfield  {author} {\bibinfo {author} {\bibfnamefont {E.}~\bibnamefont
  {Hopf}},\ }\href
  {http://www.iumj.indiana.edu/IUMJ/dfulltext.php?year=1952{\&}volume=1{\&}artid=51004}
  {\bibfield  {journal} {\bibinfo  {journal} {J. Ration. Mech. Anal.}\ }\textbf
  {\bibinfo {volume} {1}},\ \bibinfo {pages} {87} (\bibinfo {year}
  {1952})}\BibitemShut {NoStop}%
\bibitem [{\citenamefont {Wilczek}(2016)}]{Wilczek_2016}%
  \BibitemOpen
  \bibfield  {author} {\bibinfo {author} {\bibfnamefont {M.}~\bibnamefont
  {Wilczek}},\ }\href {\doibase 10.1088/1367-2630/18/12/125009} {\bibfield
  {journal} {\bibinfo  {journal} {New J. Phys.}\ }\textbf {\bibinfo {volume}
  {18}},\ \bibinfo {pages} {125009} (\bibinfo {year} {2016})}\BibitemShut
  {NoStop}%
\bibitem [{\citenamefont {Lukassen}\ and\ \citenamefont
  {Wilczek}(2017)}]{10.1007/978-3-319-57934-4_4}%
  \BibitemOpen
  \bibfield  {author} {\bibinfo {author} {\bibfnamefont {L.~J.}\ \bibnamefont
  {Lukassen}}\ and\ \bibinfo {author} {\bibfnamefont {M.}~\bibnamefont
  {Wilczek}},\ }in\ \href@noop {} {\emph {\bibinfo {booktitle} {Progress in
  Turbulence VII}}},\ \bibinfo {editor} {edited by\ \bibinfo {editor}
  {\bibfnamefont {R.}~\bibnamefont {{\"O}rl{\"u}}}, \bibinfo {editor}
  {\bibfnamefont {A.}~\bibnamefont {Talamelli}}, \bibinfo {editor}
  {\bibfnamefont {M.}~\bibnamefont {Oberlack}}, \ and\ \bibinfo {editor}
  {\bibfnamefont {J.}~\bibnamefont {Peinke}}}\ (\bibinfo  {publisher} {Springer
  International Publishing},\ \bibinfo {address} {Cham},\ \bibinfo {year}
  {2017})\ pp.\ \bibinfo {pages} {23--29}\BibitemShut {NoStop}%
\bibitem [{\citenamefont {Mardoukhi}\ \emph {et~al.}(2020)\citenamefont
  {Mardoukhi}, \citenamefont {Chechkin},\ and\ \citenamefont
  {Metzler}}]{Mardoukhi_2020}%
  \BibitemOpen
  \bibfield  {author} {\bibinfo {author} {\bibfnamefont {Y.}~\bibnamefont
  {Mardoukhi}}, \bibinfo {author} {\bibfnamefont {A.}~\bibnamefont {Chechkin}},
  \ and\ \bibinfo {author} {\bibfnamefont {R.}~\bibnamefont {Metzler}},\ }\href
  {\doibase 10.1088/1367-2630/ab950b} {\bibfield  {journal} {\bibinfo
  {journal} {New J. Phys.}\ }\textbf {\bibinfo {volume} {22}},\ \bibinfo
  {pages} {073012} (\bibinfo {year} {2020})}\BibitemShut {NoStop}%
\bibitem [{\citenamefont {Feller}(1957)}]{feller1957introduction}%
  \BibitemOpen
  \bibfield  {author} {\bibinfo {author} {\bibfnamefont {W.}~\bibnamefont
  {Feller}},\ }\href@noop {} {\emph {\bibinfo {title} {An introduction to
  probability theory and its applications}}}\ (\bibinfo  {publisher} {Wiley},\
  \bibinfo {year} {1957})\BibitemShut {NoStop}%
\bibitem [{\citenamefont {Dillon~et al.}(2017)}]{dillon2017tensorflow}%
  \BibitemOpen
  \bibfield  {author} {\bibinfo {author} {\bibfnamefont {J.}~\bibnamefont
  {Dillon~et al.}},\ }\href@noop {} {\bibfield  {journal} {\bibinfo  {journal}
  {arXiv preprint arXiv:1711.10604}\ } (\bibinfo {year} {2017})}\BibitemShut
  {NoStop}%
\bibitem [{\citenamefont {Eyink}(1993)}]{Eyink1993}%
  \BibitemOpen
  \bibfield  {author} {\bibinfo {author} {\bibfnamefont {G.~L.}\ \bibnamefont
  {Eyink}},\ }\href {\doibase http://dx.doi.org/10.1016/0375-9601(93)90117-I}
  {\bibfield  {journal} {\bibinfo  {journal} {Phys. Lett. A}\ }\textbf
  {\bibinfo {volume} {172}},\ \bibinfo {pages} {355} (\bibinfo {year}
  {1993})}\BibitemShut {NoStop}%
\bibitem [{\citenamefont {L'vov}\ and\ \citenamefont
  {Procaccia}(1996)}]{Lvov1996}%
  \BibitemOpen
  \bibfield  {author} {\bibinfo {author} {\bibfnamefont {V.}~\bibnamefont
  {L'vov}}\ and\ \bibinfo {author} {\bibfnamefont {I.}~\bibnamefont
  {Procaccia}},\ }\href {http://link.aps.org/doi/10.1103/PhysRevLett.76.2898}
  {\bibfield  {journal} {\bibinfo  {journal} {Phys. Rev. Lett.}\ }\textbf
  {\bibinfo {volume} {76}},\ \bibinfo {pages} {2898} (\bibinfo {year}
  {1996})}\BibitemShut {NoStop}%
\bibitem [{\citenamefont {R{\'e}nyi}(2007)}]{renyi2007foundations}%
  \BibitemOpen
  \bibfield  {author} {\bibinfo {author} {\bibfnamefont {A.}~\bibnamefont
  {R{\'e}nyi}},\ }\href@noop {} {\emph {\bibinfo {title} {Foundations of
  probability}}}\ (\bibinfo  {publisher} {Courier Corporation},\ \bibinfo
  {year} {2007})\BibitemShut {NoStop}%
\bibitem [{\citenamefont {Siefert}\ and\ \citenamefont
  {Peinke}(2007)}]{siefert2007complete}%
  \BibitemOpen
  \bibfield  {author} {\bibinfo {author} {\bibfnamefont {M.}~\bibnamefont
  {Siefert}}\ and\ \bibinfo {author} {\bibfnamefont {J.}~\bibnamefont
  {Peinke}},\ }\href@noop {} {\bibfield  {journal} {\bibinfo  {journal} {Phys.
  Lett. A}\ }\textbf {\bibinfo {volume} {371}},\ \bibinfo {pages} {34}
  (\bibinfo {year} {2007})}\BibitemShut {NoStop}%
\bibitem [{\citenamefont {Kullback}\ and\ \citenamefont
  {Leibler}(1951)}]{kullback1951information}%
  \BibitemOpen
  \bibfield  {author} {\bibinfo {author} {\bibfnamefont {S.}~\bibnamefont
  {Kullback}}\ and\ \bibinfo {author} {\bibfnamefont {R.~A.}\ \bibnamefont
  {Leibler}},\ }\href@noop {} {\bibfield  {journal} {\bibinfo  {journal} {Ann.
  Math. Stat.}\ }\textbf {\bibinfo {volume} {22}},\ \bibinfo {pages} {79}
  (\bibinfo {year} {1951})}\BibitemShut {NoStop}%
\bibitem [{\citenamefont {Chyzak}\ and\ \citenamefont
  {Nielsen}(2019)}]{chyzak2019closed}%
  \BibitemOpen
  \bibfield  {author} {\bibinfo {author} {\bibfnamefont {F.}~\bibnamefont
  {Chyzak}}\ and\ \bibinfo {author} {\bibfnamefont {F.}~\bibnamefont
  {Nielsen}},\ }\href@noop {} {\bibfield  {journal} {\bibinfo  {journal} {arXiv
  preprint arXiv:1905.10965}\ } (\bibinfo {year} {2019})}\BibitemShut {NoStop}%
\bibitem [{\citenamefont {Friedrich}\ \emph {et~al.}(2018)\citenamefont
  {Friedrich}, \citenamefont {Margazoglou}, \citenamefont {Biferale},\ and\
  \citenamefont {Grauer}}]{friedrich2018multiscale}%
  \BibitemOpen
  \bibfield  {author} {\bibinfo {author} {\bibfnamefont {J.}~\bibnamefont
  {Friedrich}}, \bibinfo {author} {\bibfnamefont {G.}~\bibnamefont
  {Margazoglou}}, \bibinfo {author} {\bibfnamefont {L.}~\bibnamefont
  {Biferale}}, \ and\ \bibinfo {author} {\bibfnamefont {R.}~\bibnamefont
  {Grauer}},\ }\href@noop {} {\bibfield  {journal} {\bibinfo  {journal} {Phys.
  Rev. E}\ }\textbf {\bibinfo {volume} {98}},\ \bibinfo {pages} {023104}
  (\bibinfo {year} {2018})}\BibitemShut {NoStop}%
\bibitem [{\citenamefont {De~Karman}\ and\ \citenamefont
  {Howarth}(1938)}]{de1938statistical}%
  \BibitemOpen
  \bibfield  {author} {\bibinfo {author} {\bibfnamefont {T.}~\bibnamefont
  {De~Karman}}\ and\ \bibinfo {author} {\bibfnamefont {L.}~\bibnamefont
  {Howarth}},\ }\href@noop {} {\bibfield  {journal} {\bibinfo  {journal} {Proc.
  R. Soc. A}\ }\textbf {\bibinfo {volume} {164}},\ \bibinfo {pages} {192}
  (\bibinfo {year} {1938})}\BibitemShut {NoStop}%
\bibitem [{\citenamefont {Gotoh}\ \emph {et~al.}(2002)\citenamefont {Gotoh},
  \citenamefont {Fukayama},\ and\ \citenamefont
  {Nakano}}]{gotoh-fukayama-etal:2002}%
  \BibitemOpen
  \bibfield  {author} {\bibinfo {author} {\bibfnamefont {T.}~\bibnamefont
  {Gotoh}}, \bibinfo {author} {\bibfnamefont {D.}~\bibnamefont {Fukayama}}, \
  and\ \bibinfo {author} {\bibfnamefont {T.}~\bibnamefont {Nakano}},\
  }\href@noop {} {\bibfield  {journal} {\bibinfo  {journal} {Phys. Fluids}\
  }\textbf {\bibinfo {volume} {14}},\ \bibinfo {pages} {1065} (\bibinfo {year}
  {2002})}\BibitemShut {NoStop}%
\bibitem [{\citenamefont {Shen}\ and\ \citenamefont
  {Warhaft}(2002)}]{shen2002longitudinal}%
  \BibitemOpen
  \bibfield  {author} {\bibinfo {author} {\bibfnamefont {X.}~\bibnamefont
  {Shen}}\ and\ \bibinfo {author} {\bibfnamefont {Z.}~\bibnamefont {Warhaft}},\
  }\href@noop {} {\bibfield  {journal} {\bibinfo  {journal} {Phys. Fluids}\
  }\textbf {\bibinfo {volume} {14}},\ \bibinfo {pages} {370} (\bibinfo {year}
  {2002})}\BibitemShut {NoStop}%
\bibitem [{\citenamefont {Boratav}\ and\ \citenamefont
  {Pelz}(1997)}]{boratav-pelz:1997}%
  \BibitemOpen
  \bibfield  {author} {\bibinfo {author} {\bibfnamefont {O.}~\bibnamefont
  {Boratav}}\ and\ \bibinfo {author} {\bibfnamefont {R.}~\bibnamefont {Pelz}},\
  }\href@noop {} {\bibfield  {journal} {\bibinfo  {journal} {Phys. Fluids}\
  }\textbf {\bibinfo {volume} {9}},\ \bibinfo {pages} {1400} (\bibinfo {year}
  {1997})}\BibitemShut {NoStop}%
\bibitem [{\citenamefont {Ishihara}\ \emph {et~al.}(2009)\citenamefont
  {Ishihara}, \citenamefont {Gotoh},\ and\ \citenamefont
  {Kaneda}}]{ishihara-gotoh-etal:2009}%
  \BibitemOpen
  \bibfield  {author} {\bibinfo {author} {\bibfnamefont {T.}~\bibnamefont
  {Ishihara}}, \bibinfo {author} {\bibfnamefont {T.}~\bibnamefont {Gotoh}}, \
  and\ \bibinfo {author} {\bibfnamefont {Y.}~\bibnamefont {Kaneda}},\
  }\href@noop {} {\bibfield  {journal} {\bibinfo  {journal} {Annu. Rev. Fluid
  Mech.}\ } (\bibinfo {year} {2009})}\BibitemShut {NoStop}%
\bibitem [{\citenamefont {Grauer}\ \emph {et~al.}(2012)\citenamefont {Grauer},
  \citenamefont {Homann},\ and\ \citenamefont
  {Pinton}}]{grauer-homann-pinton:2012}%
  \BibitemOpen
  \bibfield  {author} {\bibinfo {author} {\bibfnamefont {R.}~\bibnamefont
  {Grauer}}, \bibinfo {author} {\bibfnamefont {H.}~\bibnamefont {Homann}}, \
  and\ \bibinfo {author} {\bibfnamefont {J.}~\bibnamefont {Pinton}},\
  }\href@noop {} {\bibfield  {journal} {\bibinfo  {journal} {New J. Phys.}\
  }\textbf {\bibinfo {volume} {14}},\ \bibinfo {pages} {063016} (\bibinfo
  {year} {2012})}\BibitemShut {NoStop}%
\bibitem [{\citenamefont {Iyer}\ \emph {et~al.}(2017)\citenamefont {Iyer},
  \citenamefont {Sreenivasan},\ and\ \citenamefont {Yeung}}]{iyer2017reynolds}%
  \BibitemOpen
  \bibfield  {author} {\bibinfo {author} {\bibfnamefont {K.~P.}\ \bibnamefont
  {Iyer}}, \bibinfo {author} {\bibfnamefont {K.~R.}\ \bibnamefont
  {Sreenivasan}}, \ and\ \bibinfo {author} {\bibfnamefont {P.}~\bibnamefont
  {Yeung}},\ }\href@noop {} {\bibfield  {journal} {\bibinfo  {journal} {Phys.
  Rev. E}\ }\textbf {\bibinfo {volume} {95}},\ \bibinfo {pages} {021101}
  (\bibinfo {year} {2017})}\BibitemShut {NoStop}%
\bibitem [{\citenamefont {Friedrich}\ \emph {et~al.}(2016)\citenamefont
  {Friedrich}, \citenamefont {Homann}, \citenamefont {Sch{\"a}fer},\ and\
  \citenamefont {Grauer}}]{Friedrich2016}%
  \BibitemOpen
  \bibfield  {author} {\bibinfo {author} {\bibfnamefont {J.}~\bibnamefont
  {Friedrich}}, \bibinfo {author} {\bibfnamefont {H.}~\bibnamefont {Homann}},
  \bibinfo {author} {\bibfnamefont {T.}~\bibnamefont {Sch{\"a}fer}}, \ and\
  \bibinfo {author} {\bibfnamefont {R.}~\bibnamefont {Grauer}},\ }\href@noop {}
  {\bibfield  {journal} {\bibinfo  {journal} {New J. Phys.}\ }\textbf {\bibinfo
  {volume} {18}},\ \bibinfo {pages} {125008} (\bibinfo {year}
  {2016})}\BibitemShut {NoStop}%
\bibitem [{\citenamefont {Robertson}(1940)}]{robertson_1940}%
  \BibitemOpen
  \bibfield  {author} {\bibinfo {author} {\bibfnamefont {H.~P.}\ \bibnamefont
  {Robertson}},\ }\href {\doibase 10.1017/S0305004100017199} {\bibfield
  {journal} {\bibinfo  {journal} {Math. Proc. Camb. Philos. Soc.}\ }\textbf
  {\bibinfo {volume} {36}},\ \bibinfo {pages} {209} (\bibinfo {year}
  {1940})}\BibitemShut {NoStop}%
\bibitem [{\citenamefont {Chandrasekhar}(1950)}]{1950}%
  \BibitemOpen
  \bibfield  {author} {\bibinfo {author} {\bibfnamefont {S.}~\bibnamefont
  {Chandrasekhar}},\ }\href {\doibase 10.1098/rsta.1950.0010} {\bibfield
  {journal} {\bibinfo  {journal} {Phil. Trans. R. Soc. A}\ }\textbf {\bibinfo
  {volume} {242}},\ \bibinfo {pages} {557} (\bibinfo {year}
  {1950})}\BibitemShut {NoStop}%
\bibitem [{\citenamefont {Friedrich}(2020{\natexlab{a}})}]{atmos11040382}%
  \BibitemOpen
  \bibfield  {author} {\bibinfo {author} {\bibfnamefont {J.}~\bibnamefont
  {Friedrich}},\ }\href@noop {} {\bibfield  {journal} {\bibinfo  {journal}
  {Atmosphere}\ }\textbf {\bibinfo {volume} {11}} (\bibinfo {year}
  {2020}{\natexlab{a}})}\BibitemShut {NoStop}%
\bibitem [{\citenamefont {Dubrulle}(1994)}]{Dubrulle:1994ta}%
  \BibitemOpen
  \bibfield  {author} {\bibinfo {author} {\bibfnamefont {B.}~\bibnamefont
  {Dubrulle}},\ }\href {\doibase 10.1103/PhysRevLett.73.959} {\bibfield
  {journal} {\bibinfo  {journal} {Phys. Rev. Lett.}\ }\textbf {\bibinfo
  {volume} {73}},\ \bibinfo {pages} {959} (\bibinfo {year} {1994})}\BibitemShut
  {NoStop}%
\bibitem [{\citenamefont {Sosa-Correa}\ \emph {et~al.}(2019)\citenamefont
  {Sosa-Correa}, \citenamefont {Pereira}, \citenamefont {Mac{\^e}do},
  \citenamefont {Raposo}, \citenamefont {Salazar},\ and\ \citenamefont
  {Vasconcelos}}]{sosa2019emergence}%
  \BibitemOpen
  \bibfield  {author} {\bibinfo {author} {\bibfnamefont {W.}~\bibnamefont
  {Sosa-Correa}}, \bibinfo {author} {\bibfnamefont {R.}~\bibnamefont
  {Pereira}}, \bibinfo {author} {\bibfnamefont {A.}~\bibnamefont {Mac{\^e}do}},
  \bibinfo {author} {\bibfnamefont {E.}~\bibnamefont {Raposo}}, \bibinfo
  {author} {\bibfnamefont {D.}~\bibnamefont {Salazar}}, \ and\ \bibinfo
  {author} {\bibfnamefont {G.}~\bibnamefont {Vasconcelos}},\ }\href@noop {}
  {\bibfield  {journal} {\bibinfo  {journal} {Phys. Rev. Fluids}\ }\textbf
  {\bibinfo {volume} {4}},\ \bibinfo {pages} {064602} (\bibinfo {year}
  {2019})}\BibitemShut {NoStop}%
\bibitem [{\citenamefont {Sawford}(1991)}]{Sawford_1991}%
  \BibitemOpen
  \bibfield  {author} {\bibinfo {author} {\bibfnamefont {B.~L.}\ \bibnamefont
  {Sawford}},\ }\href@noop {} {\bibfield  {journal} {\bibinfo  {journal} {Phys.
  Fluids}\ }\textbf {\bibinfo {volume} {3}},\ \bibinfo {pages} {1577} (\bibinfo
  {year} {1991})}\BibitemShut {NoStop}%
\bibitem [{\citenamefont {Viggiano}\ \emph {et~al.}(2020)\citenamefont
  {Viggiano}, \citenamefont {Friedrich}, \citenamefont {Volk}, \citenamefont
  {Bourgoin}, \citenamefont {Cal},\ and\ \citenamefont
  {Chevillard}}]{Viggiano2019}%
  \BibitemOpen
  \bibfield  {author} {\bibinfo {author} {\bibfnamefont {B.}~\bibnamefont
  {Viggiano}}, \bibinfo {author} {\bibfnamefont {J.}~\bibnamefont {Friedrich}},
  \bibinfo {author} {\bibfnamefont {R.}~\bibnamefont {Volk}}, \bibinfo {author}
  {\bibfnamefont {M.}~\bibnamefont {Bourgoin}}, \bibinfo {author}
  {\bibfnamefont {R.~B.}\ \bibnamefont {Cal}}, \ and\ \bibinfo {author}
  {\bibfnamefont {L.}~\bibnamefont {Chevillard}},\ }\href
  {http://dx.doi.org/10.1017/jfm.2020.495} {\bibfield  {journal} {\bibinfo
  {journal} {J. Fluid Mech.}\ }\textbf {\bibinfo {volume} {900}} (\bibinfo
  {year} {2020})}\BibitemShut {NoStop}%
\bibitem [{\citenamefont {Friedrich}\ \emph {et~al.}(2020)\citenamefont
  {Friedrich}, \citenamefont {Gallon}, \citenamefont {Pumir},\ and\
  \citenamefont {Grauer}}]{friedrich2020stochastic}%
  \BibitemOpen
  \bibfield  {author} {\bibinfo {author} {\bibfnamefont {J.}~\bibnamefont
  {Friedrich}}, \bibinfo {author} {\bibfnamefont {S.}~\bibnamefont {Gallon}},
  \bibinfo {author} {\bibfnamefont {A.}~\bibnamefont {Pumir}}, \ and\ \bibinfo
  {author} {\bibfnamefont {R.}~\bibnamefont {Grauer}},\ }\href@noop {}
  {\bibfield  {journal} {\bibinfo  {journal} {Phys. Rev. Lett.}\ }\textbf
  {\bibinfo {volume} {125}},\ \bibinfo {pages} {170602} (\bibinfo {year}
  {2020})}\BibitemShut {NoStop}%
\bibitem [{\citenamefont {Friedrich}(2020{\natexlab{b}})}]{friedrich2020non}%
  \BibitemOpen
  \bibfield  {author} {\bibinfo {author} {\bibfnamefont {J.}~\bibnamefont
  {Friedrich}},\ }\href@noop {} {\emph {\bibinfo {title} {Non-perturbative
  Methods in Statistical Descriptions of Turbulence}}}\ (\bibinfo  {publisher}
  {Springer},\ \bibinfo {year} {2020})\BibitemShut {NoStop}%
\bibitem [{\citenamefont {{Reichherzer}}\ \emph {et~al.}(2019)\citenamefont
  {{Reichherzer}}, \citenamefont {{Becker Tjus}}, \citenamefont {{Zweibel}},
  \citenamefont {{Merten}},\ and\ \citenamefont
  {{Pueschel}}}]{reichherzer-etal:2019}%
  \BibitemOpen
  \bibfield  {author} {\bibinfo {author} {\bibfnamefont {P.}~\bibnamefont
  {{Reichherzer}}}, \bibinfo {author} {\bibfnamefont {J.}~\bibnamefont {{Becker
  Tjus}}}, \bibinfo {author} {\bibfnamefont {E.~G.}\ \bibnamefont {{Zweibel}}},
  \bibinfo {author} {\bibfnamefont {L.}~\bibnamefont {{Merten}}}, \ and\
  \bibinfo {author} {\bibfnamefont {M.~J.}\ \bibnamefont {{Pueschel}}},\
  }\href@noop {} {\bibfield  {journal} {\bibinfo  {journal} {arXiv e-prints}\
  ,\ \bibinfo {eid} {arXiv:1910.07528}} (\bibinfo {year} {2019})}\BibitemShut
  {NoStop}%
\bibitem [{\citenamefont {Giacalone}\ and\ \citenamefont
  {Jokipii}(1999)}]{giacalone-jokipii:1999}%
  \BibitemOpen
  \bibfield  {author} {\bibinfo {author} {\bibfnamefont {J.}~\bibnamefont
  {Giacalone}}\ and\ \bibinfo {author} {\bibfnamefont {J.~R.}\ \bibnamefont
  {Jokipii}},\ }\href {http://stacks.iop.org/0004-637X/520/i=1/a=204}
  {\bibfield  {journal} {\bibinfo  {journal} {ApJ}\ }\textbf {\bibinfo {volume}
  {520}},\ \bibinfo {pages} {204} (\bibinfo {year} {1999})}\BibitemShut
  {NoStop}%
\bibitem [{\citenamefont {Snodin}\ \emph {et~al.}(2016)\citenamefont {Snodin},
  \citenamefont {Shukurov}, \citenamefont {Sarson}, \citenamefont {Bushby},\
  and\ \citenamefont {Rodrigues}}]{snodin-shukurov-etal:2016}%
  \BibitemOpen
  \bibfield  {author} {\bibinfo {author} {\bibfnamefont {A.~P.}\ \bibnamefont
  {Snodin}}, \bibinfo {author} {\bibfnamefont {A.}~\bibnamefont {Shukurov}},
  \bibinfo {author} {\bibfnamefont {G.~R.}\ \bibnamefont {Sarson}}, \bibinfo
  {author} {\bibfnamefont {P.~J.}\ \bibnamefont {Bushby}}, \ and\ \bibinfo
  {author} {\bibfnamefont {L.~F.~S.}\ \bibnamefont {Rodrigues}},\ }\href@noop
  {} {\bibfield  {journal} {\bibinfo  {journal} {MNRAS}\ }\textbf {\bibinfo
  {volume} {457}},\ \bibinfo {pages} {3975} (\bibinfo {year}
  {2016})}\BibitemShut {NoStop}%
\end{thebibliography}%
\end{document}